\newcommand{\cc}{c^\dagger}

\newcommand{\be}{\begin{equation}}
\newcommand{\ee}{\end{equation}}
\newcommand{\bea}{\begin{eqnarray}}
\newcommand{\eea}{\end{eqnarray}}
\newcommand{\ben}{\begin{eqnarray*}}
\newcommand{\een}{\end{eqnarray*}}

\newcommand{\Trace}{\mathop{\rm Tr}\nolimits}

\newcommand{\averag}[1]{\langle #1 \rangle}
\newcommand{\conf}[1]{| #1 \rangle}
\newcommand{\upa}{\uparrow}
\newcommand{\doa}{\downarrow}

\newcommand{\dou}{\uparrow\downarrow}
\newcommand{\eg}{\textit{e.g.~}}
\newcommand{\ie}{\textit{i.e.~}}
\newcommand{\partder}[2]{\frac{\partial #1}{\partial #2}}

\documentclass[aps,prb,floatfix,twocolumn,showpacs,amsmath,amssymb,eqsecnum]{revtex4-1}

\usepackage{graphicx,color}
\usepackage[update,prepend]{epstopdf}
\usepackage{hyperref}
\usepackage{multirow}
\usepackage{bm}
\usepackage[T1]{fontenc}
\usepackage{float}

\begin{document}

\title{Exact solution of the 1D Hubbard model with NN and NNN
interactions in the narrow-band limit}

\author{Ferdinando Mancini$^{1,2,3}$, Evgeny Plekhanov$^{4}$, and
   Gerardo Sica$^{1,5}$}

\affiliation{\mbox{$^1$Dipartimento di Fisica ``E.R. Caianiello'',
Universit\`a degli Studi di Salerno, 84084 Fisciano (SA), Italy}\\
\mbox{$^2$Unit\`a CNISM di Salerno, Universit\`a degli Studi di Salerno,
84084 Fisciano (SA), Italy}\\
$^3$Istituto Internazionale per gli Alti Studi Scientifici (IIASS), I-84019 Vietri sul Mare (SA),
Italy\\
$^4$Consiglio Nazionale delle Ricerche (CNR-SPIN), 67100 L'Aquila, Italy\\
$^5$Department of Physics, Loughborough University, Loughborough LE11 3TU, United Kingdom}

\begin{abstract}
   We present the exact solution, obtained by means of 
   the Transfer Matrix 
   (TM) method, of the 1D Hubbard model
   with nearest-neighbor (NN) and next-nearest-neighbor (NNN) Coulomb
   interactions in the atomic limit ($t=0$).
   The competition among the
   interactions ($U$, $V_1$, and $V_2$) generates a plethora of $T=0$
   phases in the whole range of fillings. $U$, $V_1$, and $V_2$ are the
   intensities of the local, NN and NNN interactions, respectively. We
   report the $T=0$ phase diagram, in which the phases are classified
   according to the behavior of the principal correlation functions,
   and reconstruct a representative electronic configuration for each phase.
   In order to do that, we make an analytic limit $T\to 0$ in the
   transfer matrix, which allows us to obtain analytic expressions for
   the ground state energies even for extended transfer matrices.
   Such an extension of the standard TM technique can be easily applied
   to a wide class of 1D models with the interaction range beyond NN
   distance, allowing for a complete determination of the $T=0$ phase
   diagrams. 
\end{abstract}

\maketitle

\section{Introduction}

The r\^ole of the Hubbard model (HM)~\cite{hubbard_00} in the physics of
strongly correlated electronic systems can hardly be overestimated. It
has been proposed as a minimal model to explain
ferromagnetism~\cite{ferro_1,ferro_2}, stripe order~\cite{stripes_1},
paramagnetism~\cite{para_1}, metal-insulator transition~\cite{mit_1},
high-temperature superconductivity~\cite{rvb} and others. In the HM the
long-ranged Coulomb interaction is reduced to the shortest possible
distance term (on-site $U$-term).
A more realistic treatment of the Coulomb interaction should, however,
include longer-range terms.
Although the bare Coulomb interaction is repulsive at any distance,
there might be situations where some extra interactions could result in
an effective density-density interaction, thus renormalizing the Coulomb
one. It is believed~\cite{micnas_00} that if interactions with bosonic
fields, like phonons, magnons, polarons etc, are present in the system,
they would give rise to an attractive density-density term, opening the
possibility to stabilize various orders (\eg~the superconducting one). On
the other hand, the commensurate charge density orderings in several
materials, including manganites~\cite{Dagotto_00,Rao_00},
cuprates~\cite{Muschler_00}, transition metal oxides and organic
compounds~\cite{Mori_00,Seo_00}, have been argued to originate from the
nearest-neighbor (NN) and next-nearest-neighbor (NNN) density
interactions~\cite{yoshioka_00}.  Therefore, signs and magnitudes of
the long-range effective interaction terms could, in principle, vary in
a wide range.

While the HM in one dimension (1D) enjoys the integrability
property~\cite{bethe}, the addition of any extra term (\eg
density-density or spin-spin interactions at various distances) makes
the model non-integrable. In this context, any additional information
coming from an exact solution in some limiting case would be of great
help in understanding the physics of the model. In more than 1D and/or
extended with additional terms, the HM is usually studied by means of
various approximate
methods~\cite{esirgen_00,tomio_00,murakami_00,esirgen_00,su_00,su_01,
onozawa_00,tsuchiizu_00,nakamura_00,yoshioka_00,japaridze_02,vmc,giamarchi,mancini_07}.
Complementary information comes from numerical simulations of finite
systems~\cite{nakamura_01,sandvik_00,sandvik_01,ejima_00,glocke_00}. In
2D, various extended HMs (EHMs) have been proposed as key models to
explain the mechanisms for unconventional superconductivity in
cuprates~\cite{esirgen_00,su_01,su_00,onozawa_00}. The variety of
phenomena predicted in the EHMs in 2D ranges from $d$-wave
superconductivity and spin-/charge-density waves to antiferromagnetism,
ferromagnetism, and Mott metal-insulator transition at commensurate
fillings~\cite{murakami_00,esirgen_00,su_00,su_01,onozawa_00,mancini_07,mancini_18,mancini_15,mancini_16}.
However, even in 1D, the EHMs present an unexpected richness of
properties. They have been extensively analyzed in 1D by means of
bosonization
technique~\cite{tsuchiizu_00,nakamura_00,yoshioka_00,japaridze_02,giamarchi},
as well as Numerical Diagonalization~\cite{nakamura_01}, Quantum Monte
Carlo~\cite{sandvik_00,sandvik_01} and Density Matrix Renormalization
Group~\cite{ejima_00,glocke_00}. In more than 1D, EHMs have been studied
in the narrow band limit by both
exact~\cite{jedrzejewski_00,*jedrzejewski_01} and
approximate~\cite{robaszkiewicz_00} methods. Depending on the values of
Hamiltonian parameters, charge, spin and bond-order wave orderings become
relevant. Upon decreasing the kinetic energy, at some value of the
hopping parameter $t$, the system undergoes a transition towards an
insulating state. In the case of vanishing kinetic energy ($t=0$), the
electrons become frozen on the sites and interact only via the potential
energy. In the present article, we focus on such narrow-band limit.

As shown below, the HM extended by the NN and NNN density-density
interactions can be exactly solved by means of the Transfer
Matrix (TM) method~\cite{baxter} in 1D and in the narrow-band limit.
Such an EHM has been studied in the literature in distinct limiting
cases, different methods and in conjunction with various problems: as an
application to electron-lattice interaction~\cite{Bari_00}; in 1D and in
the limit $V_2=0$ by means of the TM~\cite{Beni_00,Tu_00}; Composite
Operator Method~\cite{mancini_00,mancini_33}; and within a variational
approach which treats the on-site terms exactly while treating the
inter-site ones within the mean-field
approximation~\cite{Kapcia_00,robaszkiewicz_00,Kapcia_01}.

While the application of the TM to a 1D model appears to be quite
standard, several difficulties arise due to the extended
interaction range of the model and the increased dimensionality of the
single-site Hilbert space as compared to that of the spin-$1/2$ Ising model.
The model considered here provides a concrete example with its
sixteen-dimensional TM. In addition, the interactions with a range
beyond the NN distance usually lead to a non-symmetric TM which makes
much more difficult to analyze the solution. In this paper we address
the above difficulties and, taking as an example the extended HM, show
how to symmetrize the TM, reduce the rank of the TM by using the
system's symmetry, and obtain an exact analytic $T=0$ phase diagram from
the TM matrix elements without TM diagonalization. Moreover, since the
TM results are obtained more naturally in the grand-canonical ensemble
formalism (at fixed chemical potential) we show here how to convert them
to the canonical one (at fixed particle density). Finally, by mapping
the system onto a two-level toy model at low-temperature, we analyze the
properties of the first excited state and show that $T=0$ phase
transitions occur via an interchange between the ground
and the first excited states.

The rest of this paper is organized as follows: in Sec.~\ref{modmeth} we
describe the model and introduce the TM treatment for the Hamiltonian
under investigation; in Sec.~\ref{results:phd} we present the exact
phase diagram of the model at $T=0$, while in Sec~\ref{results:therm} we
investigate the thermodynamic properties: specific heat, charge
susceptibility and entropy. Section~\ref{conclus} contains a summary of
our results and a conclusion.

\section{\label{modmeth} Model and Methods}

The Hamiltonian of the $1D$ extended Hubbard model considered in
the present manuscript is defined as follows:
\be
\begin{split}
   H &= -t\sum_{i,\sigma} \left\{ \cc_{\sigma}(i)c_{\sigma}(i+1) + \mathrm{H.c.} \right\}
   + \sum_i \left[ -\mu n(i) +U D(i) \right]\\
   &+ V_1 \sum_{i}n(i) n(i+1) + V_2 \sum_{i}n(i) n(i+2),
   \label{orig_ham}
\end{split}
\ee
where $c_{\sigma}(i)$ and $\cc_{\sigma}(i)$ are annihilation and
creation operators of electrons with spin $\sigma$ at site $i$. $\mu$ is the chemical potential, $t$
denotes the hopping between NN, while the charge density and double
occupancy at site $i$ are defined in the usual way: $n(i)=\sum_{\sigma}
n_{\sigma}(i)$, $D(i)=n_{\upa}(i)n_{\doa}(i)$, where
$n_{\sigma}(i)=\cc_{\sigma}(i)c_{\sigma}(i)$. The local, NN, and NNN
interactions are parametrized respectively by $U$, $V_1$, and $V_2$. We
measure the energy in units of $V_1$, thereby setting $|V_1|=1$. In the
present work we restrict the analysis to the narrow-bandwidth limit: $U,
V_1, V_2 \gg t$. In such a limit the Hamiltonian~(\ref{orig_ham}) takes
the form:
\be
\begin{split}
   H &= \sum_i \left[ -\mu n(i) +U D(i) \right]
   +  V_1 \sum_{i}n(i) n(i+1) \\
   &+ V_2 \sum_{i}n(i) n(i+2).
   \label{ham}
\end{split}
\ee

It is worth noting that the model under consideration can be one-to-one
mapped onto the Blume-Emery-Griffiths model~\cite{BEG} with zero
bi-quadratic interaction, first- and second-nearest neighboring
interactions, and in an external magnetic field. Indeed, by means of
the transformation 
\be
   n(i) = 1 + S(i)
\ee
where the spin variable $S(i)$ takes the values $-1,0,1$, the Hamiltonian~(\ref{ham}) transforms as
\be
   \begin{split}
	  H &= E_0 - h \sum_i S(i) + \Delta \sum_i S^2(i)\\
	  &- \sum_i 
	  \left[
	  J_1 S(i) S(i+1) + J_2 S(i) S(i+2)
	  \right]
   \label{ham_BEG}
   \end{split}
\ee
where
\be
   \begin{split}
   E_0 &= N(-\mu+V_1+V_2), \quad J_1 = -V_1, \quad J_2 = -V_2\\
   h &= \mu - \frac{1}{2}U - 2V_1 - 2V_2; \quad \Delta = \frac{1}{2} U.\\
   \end{split}
\ee

However, Hamiltonians~(\ref{ham}) and~(\ref{ham_BEG}) are not exactly
equivalent since the mapping between $S$ and $n$ should take into
account the four possible values of the particle density $n(i)$: $0,
\upa,\doa, \dou$. Letting the zero-spin state be degenerate, makes the
Hamiltonians~(\ref{ham}) and~(\ref{ham_BEG}) equivalent, provided one
redefines~\cite{mancini_32} the chemical potential $\mu$ and the on-site
potential $U$ as
\be
   \label{phr}
   \mu \to \mu - k_{B} T \ln{2}, \; U\to U - 2k_{B}T \ln{2}.
\ee

It is easy to check that after the substitution~\ref{phr} the
partition functions of the two model become identical.
 
\subsection{Transfer Matrix solution}
With the aim of applying the standard TM method, it is necessary to rewrite
the Hamiltonian in a way to contain only NN terms.
To do this we
build up ``super-sites'' each consisting of two original sites
in such a way that every super-site interacts with its neighbors
via only NN interactions.
The Hamiltonian~(\ref{ham}) can be written as:
\be
\begin{split}
   H &= \sum_{i=1}^{2N} \left[ -\mu n(i) +U D(i) \right]
   + V_1 \sum_{i=1}^{2N} n(i) n(i+1) \\
  &+ V_2 \sum_{i=1}^{2N} n(i) n(i+2).
\end{split}
   \label{hamii}
\ee
Hereafter the periodic boundary conditions (PBC) with $2N$ sites are assumed.
We rewrite~(\ref{hamii}) in order to underline the subdivision into odd and even
sites:
\begin{widetext}
   \be
   \begin{split}
   H &= \sum_{k=1}^{N} \left[ -\mu n(2k  ) +U D(2k  ) \right]
      + \sum_{k=1}^{N} \left[ -\mu n(2k-1) +U D(2k-1) \right] 
      + V_1 \sum_{k=1}^{N} n(2k) n(2k+1)\\
     &+ V_1 \sum_{k=1}^{N} n(2k-1) n(2k)
      + V_2 \sum_{k=1}^{N} n(2k) n(2k+2)
      + V_2 \sum_{k=1}^{N} n(2k-1) n(2k+1).
   \end{split}
   \label{longtm}
   \ee
\end{widetext}
A super-site number $k$ consists of two original sites $2k-1$ and $2k$
(whose occupation numbers we will refer in what follows as $n_1(k)$ and
$n_2(k)$) respectively and their NN interaction $V_1 \sum_{k=1}^{N}
n(2k-1) n(2k)$ together with their on-site terms parametrized by $\mu$
and $U$. The on-site super-site Hamiltonian part reads as:
\be
\begin{split}
   &S_k = -\mu \left[ n_1(k) + n_2(k) \right]\\
   &+ U \left[ D_1(k) + D_2(k) \right]
    + V_1  n_1(k) n_2(k).
   \label{intrs}
\end{split}
\ee
The inter-site super-site Hamiltonian part amounts to:
\be
\begin{split}
   &P_{k,k+1} = V_1 n_2(k) n_1(k+1) \\
   &+ V_2 \left\{ n_1(k) n_1(k+1)
   + n_2(k) n_2(k+1) \right\}.
   \label{isss}
\end{split}
\ee
The whole Hamiltonian~(\ref{longtm}) can be rewritten as follows:
\[
H = \sum_{k=1}^{N} \left\{ P_{k,k+1} + S_k \right\}=
   \sum_{k=1}^{N} \left[ P_{k,k+1} + \frac{S_k+S_{k+1}}{2} \right].
\]
Following the Transfer Matrix~(TM) method, we write down the
partition function:
\be
\begin{split}
   Z &= \Trace \exp\left\{ -\beta\sum_{k=1}^{N} \left[ P_{k,k+1} +
   \frac{S_k+S_{k+1}}{2} \right] \right\}\\
   &=\prod_{k=1}^{N} Z_1(k,k+1) =Z_1^N(1,2).
\end{split}
\ee
Here $Z_1(1,2)=\exp\left\{ -\beta \left[ P_{1,2}+\frac{S_1+S_{2}}{2}
\right] \right\}$ is a matrix defined in coordinates of the first and
second super-sites: each matrix element of $Z_1(1,2)$ is given by:
\be
   \label{trans_mat}
   Z_1(1,2)_{i,j} = \exp\left\{ -\beta K_{i,j} \right\},
\ee
where $K_{i,j}=\left[ P_{1,2}+\frac{S_1+S_{2}}{2} \right]_{i,j}$ and the
basis set for rows and columns of $K_{i,j}$ is defined in
Table~\ref{basis}.
As usual, in the thermodynamic limit the free energy can be calculated
by means of the maximum eigenvalue $\lambda_{\max}$ of $Z_1(1,2)$:
\be
   \frac{1}{N}\lim_{N\to\infty} F=-T\log{\lambda_{\max}}.
\ee

A non-symmetric $16\times16$ matrix $Z_1(1,2)_{i,j}$ can be easily
diagonalized numerically by using a standard diagonalization routines
library (\eg LAPACK) for each choice of the Hamiltonian parameters and
temperature. It is much easier, however, to deal with a symmetric matrix
for what concerns the numerical diagonalization. In our case the
non-symmetry of the TM $Z_1(1,2)$ arises from the first term
proportional to $V_1$ in~(\ref{isss}). One can easily see how the TM can
be symmetrized without changing its eigenvalues. Indeed, from one hand
we have PBC applied, from the other hand we split the system into the
super-sites starting from site number $1$. We can do the same splitting
starting, say from site $2$, which is equivalent to the following index
shift in Eqs.~(\ref{intrs}-\ref{isss}):
\bea
   n_1(k) &\to& n_2(k)\\
   n_2(k) &\to& n_1(k+1).
\eea
Under such a transformation, $S_k$ transforms into itself, while in
$P_{k,k+1}$ the first term becomes:
\be
   V_1 \sum_{k=1}^{N} n_1(k) n_2(k+1).
\ee
The newly transformed Hamiltonian is exactly the same as the original
one, the difference is merely due to a change of notation,
allowed by the translational invariance of the system. We use
the sum of the two representations divided by two as an equivalent form
for the Hamiltonian leading to a symmetric TM. In this
equivalent form, $S_k$ is given by~(\ref{intrs}), while:
\[
\begin{split}
   &P_{k,k+1} = \frac{V_1}{2} 
   \left\{ n_1(k) n_2(k+1) + n_2(k) n_1(k+1)\right\}\\
   &+ V_2 
   \left\{ n_1(k) n_1(k+1) + n_2(k) n_2(k+1) \right\}.
\end{split}
\]
\section{\label{results:phd} $T=0$ phase diagram}
\begin{table}
   \caption{\label{basis} Basis set of the super-site used to define the TM.}
   \begin{ruledtabular}
   \begin{tabular}{rlrlrlrl}
   1) &$\conf{0,0}$    &  2)&$\conf{0,\upa}$    &  3)&$\conf{0,\doa}$    &  4)&$\conf{0,\dou}$   \\
   5) &$\conf{\upa,0}$ &  6)&$\conf{\upa,\upa}$ &  7)&$\conf{\upa,\doa}$ &  8)&$\conf{\upa,\dou}$ \\
   9) &$\conf{\doa,0}$ & 10)&$\conf{\doa,\upa}$ & 11)&$\conf{\doa,\doa}$ & 12)&$\conf{\doa,\dou}$ \\
   13)&$\conf{\dou,0}$ & 14)&$\conf{\dou,\upa}$ & 15)&$\conf{\dou,\doa}$ & 16)&$\conf{\dou,\dou}$
\end{tabular}
\end{ruledtabular}
\end{table}
At very low temperature, the elements of the TM start either to diverge
or tend to zero due to exponential dependence on temperature. Each
matrix element of $Z_1(1,2)$ has the form $\exp\{ -E_{i,j}/T\}$. We call
the quantity $E_{i,j}$ - energy scale associated with the matrix element
$Z_1(1,2)_{i,j}$. The lowest energy scale of the TM determines the $T\to
0$ limit of the free energy. It is possible to find an analytical limit
of the free energy at $T\to 0$ in the model~(\ref{ham}).
In this section we find the $T\to 0$ limit of the free energy
for the Hamiltonian~(\ref{ham}) in the whole range of its parameters;
then, by passing from the free energy to the internal one, we construct
the complete phase diagram of the model and characterize the properties
of each phase. The approach presented here is quite general and can be
applied to any system, tractable with a finite-dimensional TM in 1D.
\begin{table}
\begin{ruledtabular}
   \caption{\label{en_scales}
   All independent energy scales of the $16\times16$ TM
   given by the formula~(\ref{trans_mat}).}
   \begin{tabular}{l|c}
   $n$ & Energy scale \\
   \hline
   \hline
   $0$ & $0$ \\
   \hline
   $\frac{1}{4}$ & $-\frac{\mu}{4}$ \\[2pt]
   \hline
   $\frac{1}{2}$ & $\frac{V_2}{2}-\frac{\mu}{2}; \quad \frac{U}{4}-\frac{\mu}{2}; \quad \frac{V_1}{4}-\frac{\mu}{2}$\\[2pt]
   \hline
   $\frac{3}{4}$ & $\frac{U}{4}+V_2-\frac{3}{4}\mu; \quad \frac{U}{4}+\frac{V_1}{2}-\frac{3}{4}\mu; \quad
                     \frac{1}{2}(V_1+V_2)-\frac{3}{4}\mu$\\[2pt]
   \hline
   \multirow{2}{*}{$1$} & $\frac{U}{2}+2 V_2-\mu; \quad \frac{U}{2}+ V_1-\mu; \quad \frac{U}{4}+\frac{3}{4} V_1 + V_2-\mu$\\
	   & $\frac{U}{4}+V_1+ \frac{V_2}{2}-\mu; \quad V_1+ V_2-\mu$ \\[2pt]
   \hline
   \multirow{2}{*}{$\frac{5}{4}$} &  $\frac{U}{2}+V_1+2 V_2 - \frac{5}{4}\mu; \quad \frac{U}{2}+\frac{3}{2}V_1+ V_2 - \frac{5}{4}\mu$\\
                 &  $\frac{U}{4}+\frac{3}{2}( V_1+ V_2 ) - \frac{5}{4}\mu$ \\[2pt]
   \hline
   \multirow{2}{*}{$\frac{3}{2}$} &  $\frac{U}{2}+ 2V_1 +\frac{5}{2} V_2-\frac{3}{2}\mu; \quad \frac{3}{4}U + 2V_1 + 2V_2 -\frac{3}{2}\mu$\\ 
                 & $\frac{U}{2} + \frac{9}{4} V_1 + 2V_2-\frac{3}{2}\mu$\\[2pt]
   \hline
   $\frac{7}{4}$ & $\frac{3}{4}U + 3V_1 + 3V_2-\frac{7}{4}\mu$ \\[2pt]
   \hline
   $2$ & $U+4V_1+4V_2 - 2\mu$ \\
\end{tabular}
\end{ruledtabular}
\end{table}

In our case, the TM $Z_1(1,2)$ is a $16\times16$ matrix reported in
Appendix~\ref{app:a}; however a further reduction is possible. Indeed, by inspecting the
super-site basis states listed in Table~\ref{basis} and taking into account that the
Hamiltonian~(\ref{ham}) depends only on the total density on a given site, we conclude
that the following rows and columns of $Z_1(1,2)$ are identical (\ie for all $i$ the following
relation holds):
\[
\begin{array}{lclllcl}
  Z_1(1,2)_{2i} &=& Z_1(1,2)_{3i} ;&\phantom{=}& Z_1(1,2)_{i2}  &=& Z_1(1,2)_{i3} \\
  Z_1(1,2)_{5i} &=& Z_1(1,2)_{9i} ;&\phantom{=}& Z_1(1,2)_{i5}  &=& Z_1(1,2)_{i9} \\
  Z_1(1,2)_{8i} &=& Z_1(1,2)_{12i};&\phantom{=}& Z_1(1,2)_{i8}  &=& Z_1(1,2)_{i12}\\
  Z_1(1,2)_{14i}&=& Z_1(1,2)_{15i};&\phantom{=}& Z_1(1,2)_{i14} &=& Z_1(1,2)_{i15}\\
  Z_1(1,2)_{6i} &=& Z_1(1,2)_{7i}  &=& Z_1(1,2)_{10i} &=& Z_1(1,2)_{11i} \\
  Z_1(1,2)_{i6} &=& Z_1(1,2)_{i7}  &=& Z_1(1,2)_{i10} &=& Z_1(1,2)_{i11}.
\end{array}
\]
This means that the actual rank of the matrix is at most $9$, while
at least seven roots of the characteristic polynomial of $Z_1(1,2)$ are
zero.

Furthermore, it can be easily seen that the Hamiltonian~(\ref{ham}) is
invariant with respect to the simultaneous interchange of the sites
inside the super-sites corresponding to the raw and column states of
$Z_1(1,2)$. Such an interchange defines a $Z_2$ Abelian symmetry group
and connects the following couples of states from Table~\ref{basis}: $\#2 \leftrightarrow \#5$,
$\#4 \leftrightarrow \#13$ and $\#8 \leftrightarrow \#14$. One can prove that
this symmetry operation introduces a linear dependence among the six states mentioned above so that
only four of them are linearly independent.
This further reduces the rank of $Z_1(1,2)$ to $7$. The aim of such a
rank reduction procedure is to identify all the independent exponents in the
TM matrix elements, since they define the energy scales of our system.
The direct count of the matrix elements of symmetry reduced TM results in
21 independent energy scales.

It is clear now how to obtain the $T=0$ phase diagram of the model: for each energy scale
we can find a set of inequalities determining the region of the Hamiltonian
parameters where this energy scale is the lowest. In doing that, it is
convenient to classify the energy scales in groups based on the values of $n$
they correspond to, as shown in Table~\ref{en_scales}. For given values of $U$,
$V_1$, $V_2$ and $\mu$, inside each group, the group minimum can be easily
determined. The global minimum is chosen among the group minima. Since we fix
$\mu$, while $n$ is determined by requesting the global minimum of the free
energy, we work in the grand canonical ensemble. As often happens in
narrow-band models, by varying $\mu$, $n$ can only assume values from a
discrete set $n_{comm}$, as evidenced in the first column of the
Table~\ref{en_scales}, where
$n_{comm}=0,\frac{1}{4},\frac{1}{2},\frac{3}{4},1,\frac{5}{4},\frac{3}{2},\frac{7}{4},2$.
One can note that the number of possible energy scales is maximal when $n=1$ and
the energy scales obey the particle-hole symmetry relations:
$F_i(2-n)-F_i(n)=(1-n)(A-2\mu)$ for any $n\in n_{comm}\ne 1$ and $i$ inside the group
of energy scales corresponding to $n$. Here the auxiliary quantity $A$ is
defined as: $A=U+4(V_1+V_2)$. Another consequence of the particle-hole symmetry
is the relation between the values of the chemical potential on different sides of half
filling: $\mu(2-n)+\mu(n)=A$. We can, therefore, restrict our description of the
$T=0$ phase diagram to the case $0\leqslant n\leqslant 1$. In such a way, it is
possible to obtain the free energy $F(\mu)$ and the occupation per site $n(\mu)$
as functions of chemical potential in the whole range of $U$, $V_1$ and $V_2$.

Let us analyze $n(\mu)$ in details. $n(\mu)$ is a constant function with a
series of jumps at some values $\nu_{i}$ of the chemical potential.
At the jumps, $n$ changes among the values belonging to $n_{comm}$.
The criterion to determine the jumps of $n(\mu)$ follows from the
requirement of stability of the system, namely that $n$ be an increasing
function of $\mu$. In order to analyze the Hamiltonian~(\ref{ham}), it is
extremely useful to introduce the following quantities:
\be
\begin{split}
   \mu_1^{\phantom{\star}} &= \min\{V_1,U,2V_2\}\\
   \mu_3^{\phantom{\star}} &= \left \{
   \begin{array}{ll}
	  U+4V_2    , & 2V_2<V_1, \; U<2V_1-2V_2\\
	  U+2V_1    , & 2V_2>V_1, \; U<2V_2\\
	  2(V_1+V_2), & U>  2V_2, \; U>2V_1-2V_2.
   \end{array}
   \right.
\end{split}
\ee
%
The advantage of using the quantities $\mu_1$,$\mu_3$ consists in
the possibility to adopt the following compact notation for the
dependence $F(\mu)$ at various values of $n$:
\be
   \label{fofn}
   F(\mu) = \left\{
   \begin{array}{ll}
   0,           & n=0 \\
   -\frac{\mu}{4}, & n=\frac{1}{4} \\
   \frac{\mu_1}{4} -\frac{\mu}{2}, & n=\frac{1}{2}  \\
   \frac{\mu_3}{4} -\frac{3}{4}\mu, & n=\frac{3}{4} \\
   \frac{\mu_3}{2} -\mu, & n=1.
   \end{array}
   \right.
\ee
Here the minimization of the free energy inside each group of the energy scales
corresponding to a determinate value of $n$ has already been
accomplished.
At any given $\mu$, the free energy should be minimal.
Therefore, the average particle number per site is determined by
the corresponding minimal energy scale.
Thus, we have the
following upper and lower bounds for $\mu$ in various states at fixed
$n$:
\be
   \label{in_ineq}
   \begin{array}{lll}
	  n=0&:           &         \mu < \nu_1 \\
	  n=\frac{1}{4}&: & 0     < \mu < \nu_2 \\
	  n=\frac{1}{2}&: & \nu_3 < \mu < \mu_3 - \nu_3 \\
	  n=\frac{3}{4}&: & \mu_3-\nu_2 < \mu < \mu_3 \\
	  n=1&:           & \mu_3-\nu_1 < \mu < A - \mu_3 + \nu_1,
   \end{array}
\ee
where we have introduce the quantities $\nu_1-\nu_3$:
\be
   \label{nus}
   \begin{split}
	  \nu_1 &= \min\left( 0, \frac{\mu_1}{2}, \frac{\mu_3}{3},
	  \frac{\mu_3}{2}\right) \\
	  \nu_2 &= \min\left( \mu_1, \frac{\mu_3}{2},
	  \frac{2\mu_3}{3}\right) \\
	  \nu_3 &= \max\left(\frac{\mu_1}{2}, \mu_1 \right).
   \end{split}
\ee
If the upper bound for $\mu$ goes below the lower bound at a
given $n$, the state becomes thermodynamically unstable and
disappears. It can be easily seen from~(\ref{in_ineq})-(\ref{nus}) that
the existence conditions for $n=\frac{1}{4}$ and $n=\frac{3}{4}$
coincide. Considering the number of jumps in the dependence
$n(\mu)$ in the range $n\in [0,1]$ we can distinguish four cases:
i) four jumps at $n=\frac{1}{4}, \frac{1}{2}, \frac{3}{4}, 1$; ii) two
jumps at $n=\frac{1}{2}, 1$; iii) one jump at $n=1$; iv) zero jumps. In
the latter case, the only jump in the whole dependence $\mu(n)$ occurs at
$n=2$. We have analyzed the inequalities in each of the
cases and reconstructed the dependence $n(\mu)$ in the case of arbitrary
$U$, $V_1$, $V_2$ and $\mu$. At this point we need a procedure to convert our
results to the canonical ensemble.

\begin{figure*}
   \includegraphics[width=4.25cm,angle=270]{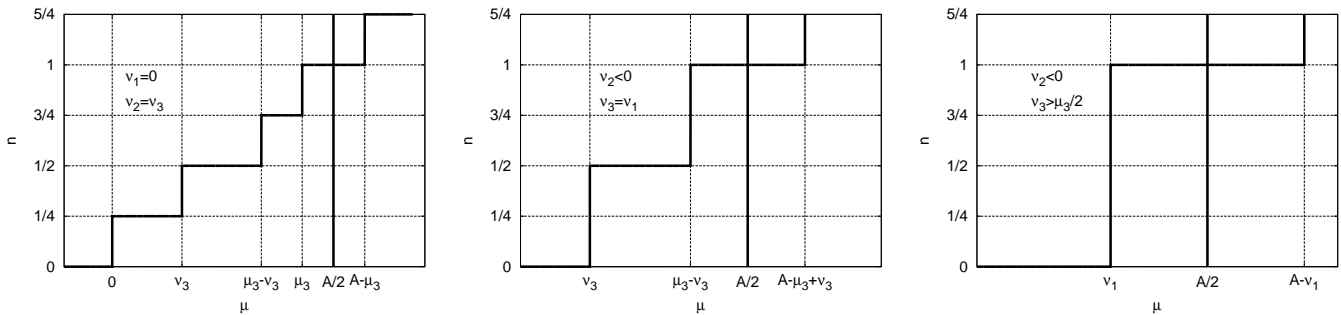}
   \caption{The dependence $\mu(n)$ in three different cases, depending
   on the number of jumps in the range $[-\infty,\frac{A}{2}]$:
   four jumps (left panel);
   two jumps (middle panel); one jump (right panel).}
   \label{fig1}
\end{figure*}

In order to pass to the canonical ensemble (fixed $n$, while varying $\mu$), we
have to invert the dependence $n(\mu)$ and pass to $\mu(n)$: so that $F(\mu)\to
F(\mu(n))\to F(n)$. At zero temperature, such a dependence is, in general, a
function with a series of steps (the consequence of finiteness of $n_{comm}$),
as shown in Fig.~\ref{fig1}. The step function does not allow the inversion
because of the undefined function and derivative values at jumps. However, we
can imagine a process of where the system is cooled adiabatically, so that each time we
deal with a system at finite $T$, while $n(\mu)$ remains a well-defined and
differentiable function, including the jumps, and therefore the inversion is
well defined. In addition, it can be easily seen that the free energy is a
linear function of $\mu$ and $n$. Taking into account the above considerations,
the inversion $n(\mu)\to\mu(n)$ becomes obvious and we briefly sketch below how
to convert $F(\mu)$ to $F(n)$. Such a conversion is accomplished piecewise and
here we illustrate it in the range between $n=n_1$ and $n=n_2$, both belonging
to $n_{comm}$. Suppose that the jump of $n(\mu)$ from $n_1$ and $n_2$ occurs at
$\mu=\mu^{\star}$. Since $F(n)$ is a linear function and it is known in two
points $F_1=F(\mu^{\star-})$ and $F_2=F(\mu^{\star+})$ the whole function in the
range $n\in[n_1,n_2]$ can be fixed. Once the dependence $F(n)$ is determined in
the whole range of $n$, we can easily determine the internal energy:
$E(n)=F(n)+\mu(n) n$.
%
In this way an
analytic expression for $E(n)$ can be obtained for arbitrary values of
the Hamiltonian parameters. By comparing $E(n)$ with~(\ref{ham}) we can
infer the expressions for the following CFs which we call {\it fundamental}:
\be
   \begin{split}
	  \partder{E}{U}   &= \averag{n_{\upa}(i)n_{\doa}(i)} = D\\
	  \partder{E}{V_1} &= \averag{n(i)n(i+1)} = C_1\\
	  \partder{E}{V_2} &= \averag{n(i)n(i+2)} = C_2.
   \end{split}
   \label{fund_cfs}
\ee
These fundamental CFs completely describe the ground-state and allow to
reconstruct the typical density pattern characterizing a given phase.

%

At $T=0$, two types of phases can be distinguished. The phases of the
first type, which we call \textit{commensurate phases}, occur at the
commensurate fillings $n_{comm}$ and in certain range of the Hamiltonian
parameters, provided that $\mu(n)$ has a jump at those fillings. In the
commensurate phases, the free energy is determined directly by one of
the $21$ energy scales. The electronic density arrangement is
characterized by periodic patterns with a finite degeneracy in the
thermodynamic limit; such patterns can be traced back to the super-sites
configurations corresponding to the energy scale realizing the global
minimum. In addition, in these phases, the chemical potential is not
fixed by $n$, but can vary in a certain range (the consequence of the
jump).

The second phase type is realized at the incommensurate fillings, or even at the
commensurate ones when there are no jumps in $\mu(n)$ at those fillings. In these
phases, the chemical potential remains constant as $n$ increases.
Since the number of competing interactions ($U$, $V_1$ and $V_2$)
is considerable, a large amount of phases of this type emerges in the phase
diagram of the system upon varying the filling. A careful analysis of the whole
phase space of the model~(\ref{ham}) reveals as many as $20$ second type phases
at $V_1>0$ and $8$ ones at $V_1<0$. In each of these phases, the fundamental
CFs~(\ref{fund_cfs}) are known analytically and density patterns representing
the ground state electron configurations can be easily reconstructed. These
analytic expressions, together with the inequalities determining the phase
boundaries, are reported in Appendix~\ref{app:e}. Given the large number of
the second type phases, in the present manuscript, we group together several of
them depending on the criterion described below. We call such phase groups
\textit{macro-phases}. All the phases belonging to the same macro-phase have the
same fundamental CFs different from zero. Such criterion simplifies
significantly the phase diagram landscape, still maintaining the description
physically meaningful. Indeed, phases which have \eg $C_1\ne 0$, $C_2 \ne 0$ and
$D=0$ are ``similar'' in the sense that they have the density correlations at NN
and NNN neighbor distance and do not have double occupancy, while the actual
functional form of CFs may be different in different phases.

We start now the description of the phases starting from the commensurate ones and then
proceeding with the incommensurate macro-phases. Since $|V_1|$ is used as the energy unit, two cases
can be distinguished: $V_1=+1$ and $V_1=-1$.
\subsection{Case $V_1=+1$}
\begin{figure*}
   \centering
   \includegraphics[width=4.2cm,angle=270]{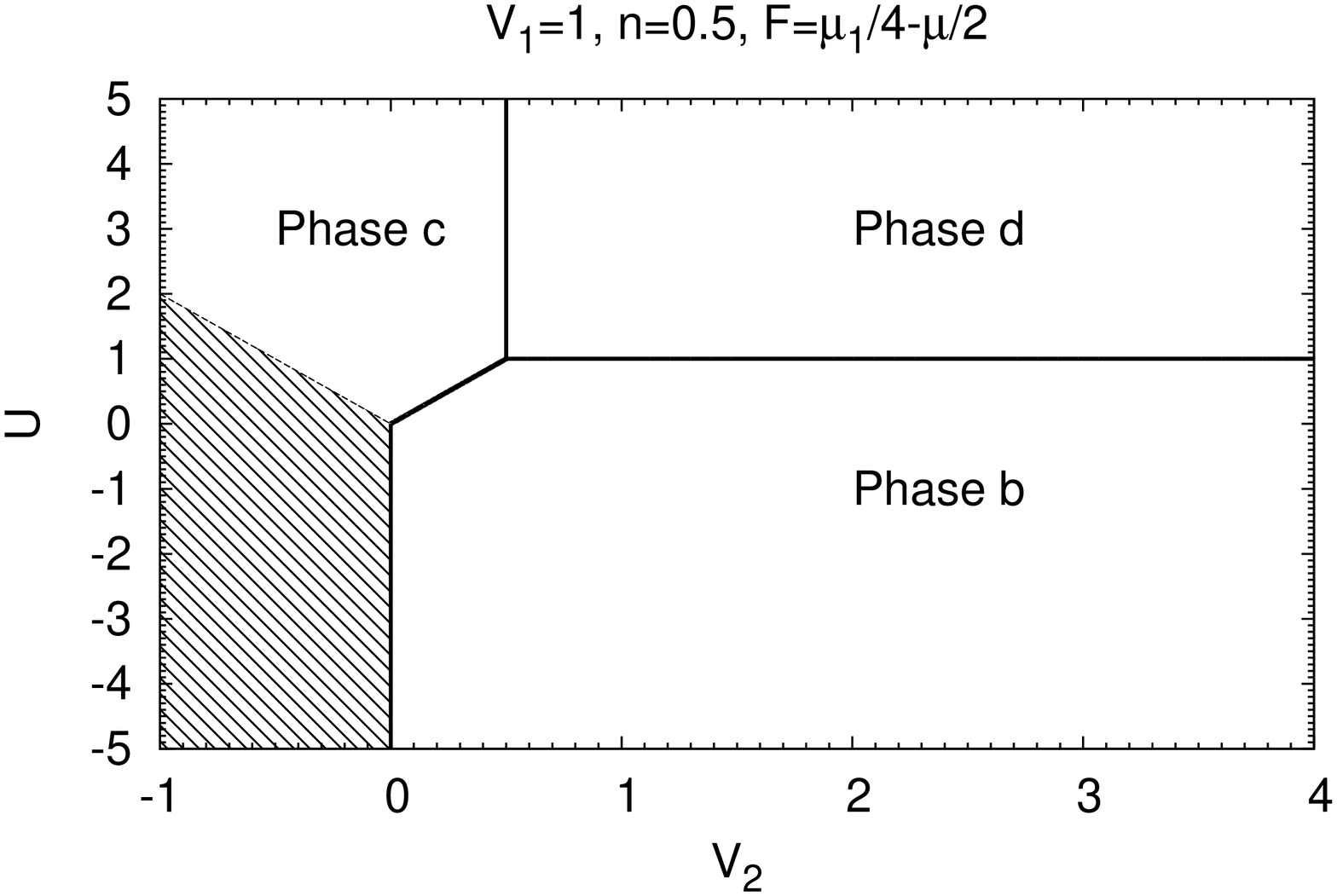}
   \includegraphics[width=4.2cm,angle=270]{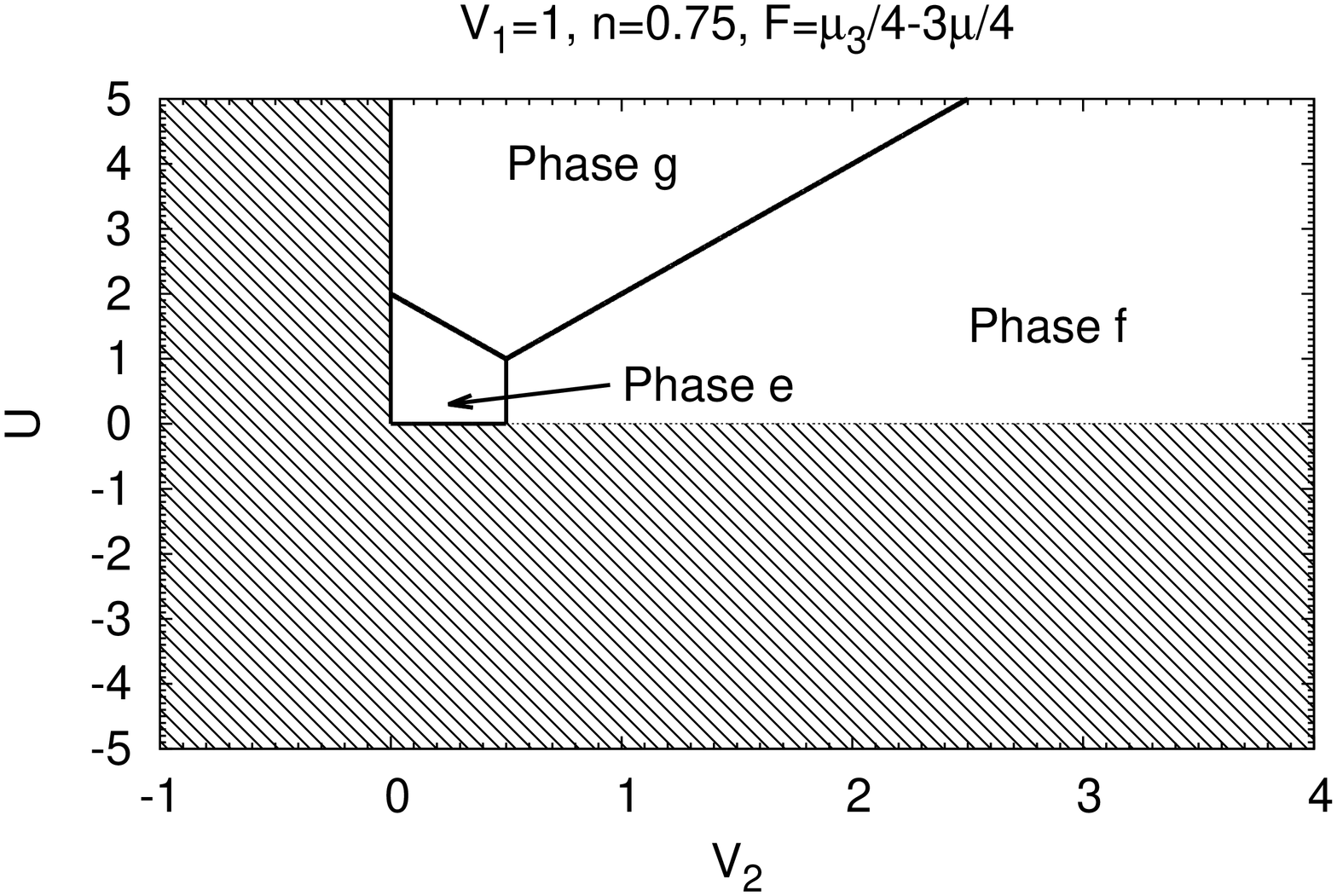}
   \includegraphics[width=4.2cm,angle=270]{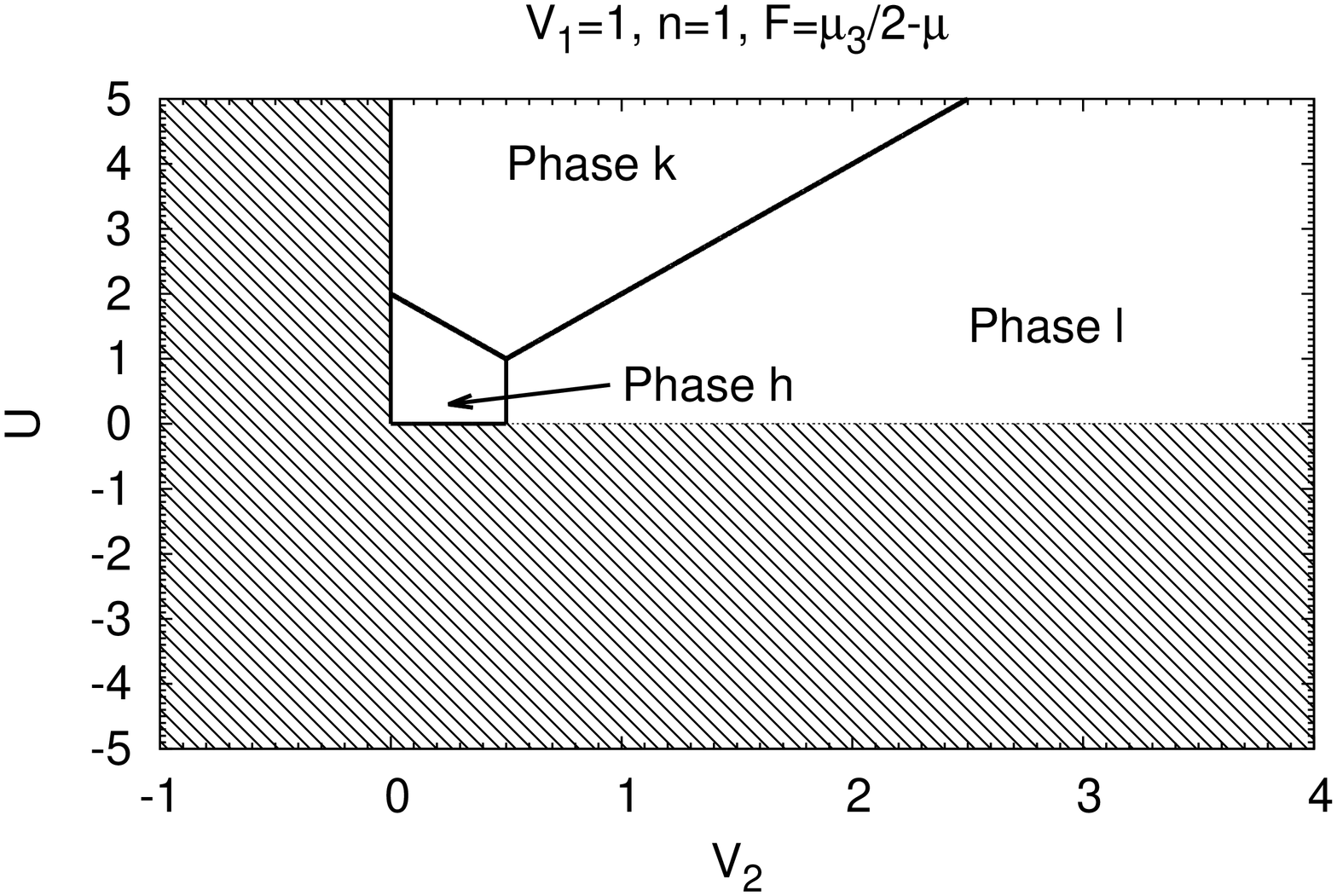}
   \caption{Phase diagram in the case $V_1=1$ for $n=0.5$ (left panel),
   $n=0.75$ (middle panel), and $n=1$ (right panel), taking into
   consideration the constrains $0<\mu_1<\mu_2<\mu_3<\mu_4$. In the
   shaded areas the jump in the dependence
   $\mu(n)$ is absent at the given commensurate $n$.}
   \label{fig2}
\end{figure*}

The commensurate phases in the case $V_1=+1$ can exist at
$n_{comm}=0,\frac{1}{4},\frac{1}{2},\frac{3}{4},1$.
We denote them by the latin letters from $a$ to $l$.
The \textit{tricritical point} (the critical point which separates three
phases) exists for all values of $n_{comm}$ at $V_2=\frac{V_1}{2}$, $U=V_1$.
We skip the trivial phase at $n=0$ and start from $n=\frac{1}{4}$.
The commensurate phase diagrams in this case are depicted in
Fig.~\ref{fig2}.
\subsubsection{$n=\frac{1}{4}$}
\emph{Phase a:} is characterized by a low density of electrons, in fact,
none of the interaction terms is working in this phase.
Singly occupied sites are separated by at least two empty ones. The chemical potential:
$0<\mu<\min\{V_1,U,2V_2\}$. All the fundamental CFs are zero. This
phase exists in the range $U>0\land V_2>0$.
\subsubsection{$n=\frac{1}{2}$}
\emph{Phase b:} is characterized by the presence of only doubly occupied sites separated by
three empty ones. A typical occupation pattern can be represented as: $\left|D,0,0,0\right\rangle$.
This phase exists in the range $U<2V_2 \land U<V_1 \land 0<V_2$. Energy per site: $\frac{U}{4}$. Chemical
potential: $U<\mu< 2\min\{V_1,2V_2\}$.

\emph{Phase c:} is characterized by the presence of only singly occupied sites at NN
distance and interacting via $V_2$. A typical occupation pattern can be represented as:
$\left|\sigma,0,\sigma^{\prime},0\right\rangle$.
This phase exists in the range $U>2V_2 \land U>-2V_2 \land 2V_2<V_1$.
Energy per site: $\frac{V_2}{2}$. Chemical potential: $2V_2< \mu< \min\{2V_1,U+2V_2\}$.

\emph{Phase d:} is characterized by the presence of only singly occupied
sites residing on NN sites and interacting via $V_1$. A typical occupation pattern can be represented as:
$\left|\sigma,\sigma^{\prime},0,0\right\rangle$. 
This phase exists in the range $U>1 \land 2V_2>V_1$. Energy per site:
$\frac{V_1}{4}$. Chemical potential: $V_1< \mu< V_1+\min\{U,2V_2\}$.
\subsubsection{$n=\frac{3}{4}$}
\emph{Phase e:} is characterized by the following pattern:
$\conf{0,\sigma,0,D}$. This phase exists in the range $0<2V_2<V_1 \land 0<U<2V_1-2V_2$.
Energy per site: $\frac{U}{4}+V_2$. Chemical potential:
$2V_2+\max\{U, 2V_2\}<\mu< U+4V_2$.

\emph{Phase f:} is characterized by the following pattern
$\conf{0,\sigma,D,0}$. This phase exists in the range $2V_2>V_1 \land 0<U<2V_2$.
Energy per site: $\frac{U}{4}+\frac{V_1}{2}$. Chemical potential:
$V_1+\max\{V_1,U\}< \mu< U+2V_1$.

\emph{Phase g:} is characterized by the following pattern
$\conf{0,\sigma,\sigma^{\prime},\sigma^{\prime\prime}}$.
This phase exists in the range $0<2V_2<U \land
U>2V_1-2V_2$.
Energy per site: $\frac{1}{2}(V_1+V_2)$. Chemical potential:
$V_1+\max\{V_1,2V_2\}< \mu< 2(V_1+V_2)$.
\subsubsection{$n=1$}
\emph{Phase h:} is characterized by the following pattern
$\conf{0,D,0,D}$. This phase exists in the range $0<2V_2<V_1 \land 0<U<2V_1-2V_2$.
Energy per site: $\frac{U}{2}+2V_2$. Chemical potential: $U+4V_2<\mu<4V_1$.

\emph{Phase k:} is characterized by the following pattern
$\conf{0,0,D,D}$. This phase exists in the range $2V_2>V_1 \land 0<U<2V_2$.
Energy per site: $\frac{U}{2}+V_1$. Chemical potential: $2(V_1+V_2)<\mu<U+2(V_1+V_2)$.

\emph{Phase l:} is characterized by the following pattern
$\conf{\sigma,\sigma^{\prime},\sigma^{\prime\prime},\sigma^{\prime\prime\prime}}$.
This phase exists in the range $0<2V_2<U
\land U>2V_1-2V_2$. Energy per site: $V_1+V_2$. Chemical potential: $U+2V_1<\mu<2V_1+4V_2$.

\begin{figure*}
   \centering
   \includegraphics[width=4.2cm,angle=270]{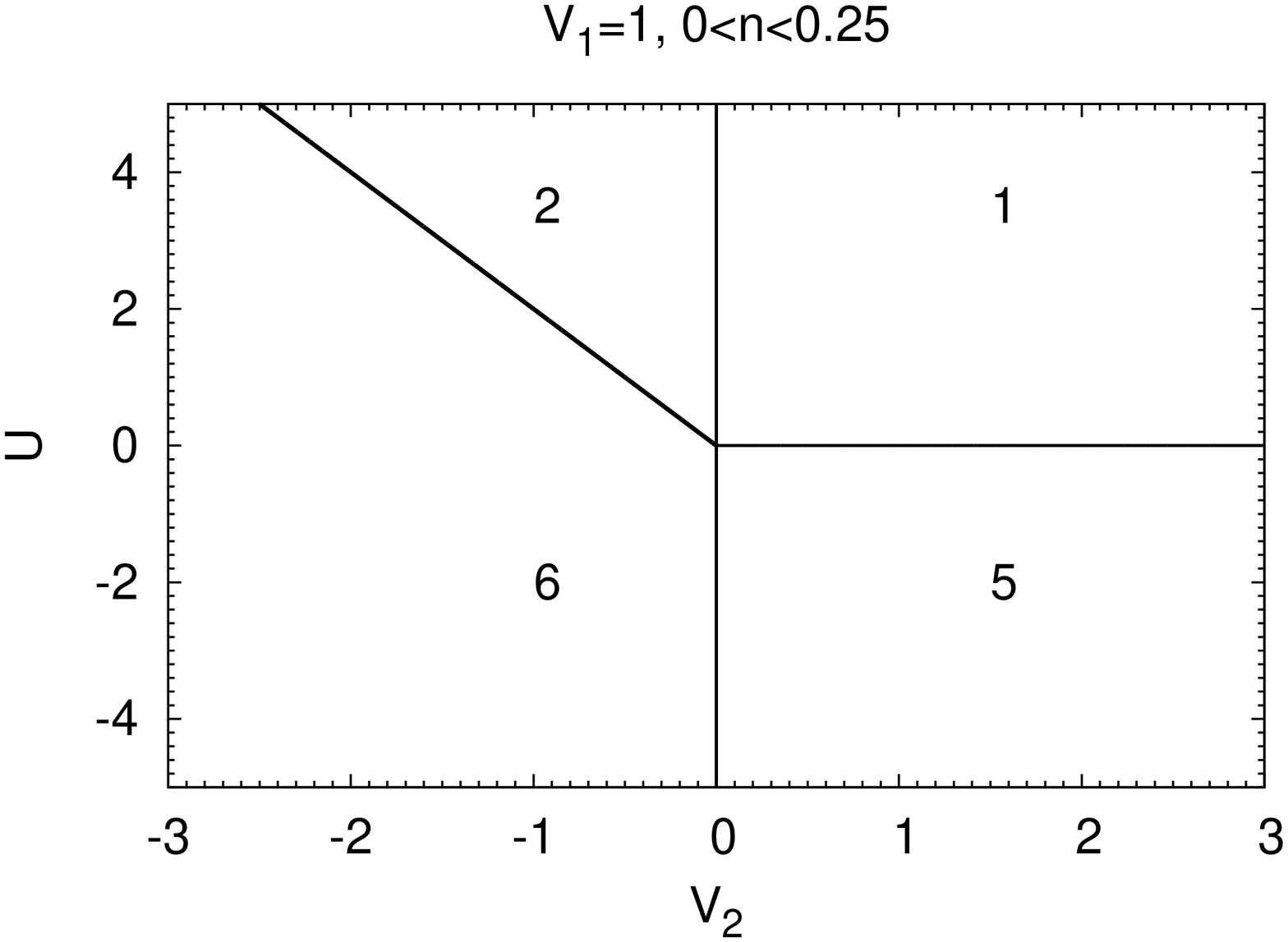}
   \includegraphics[width=4.2cm,angle=270]{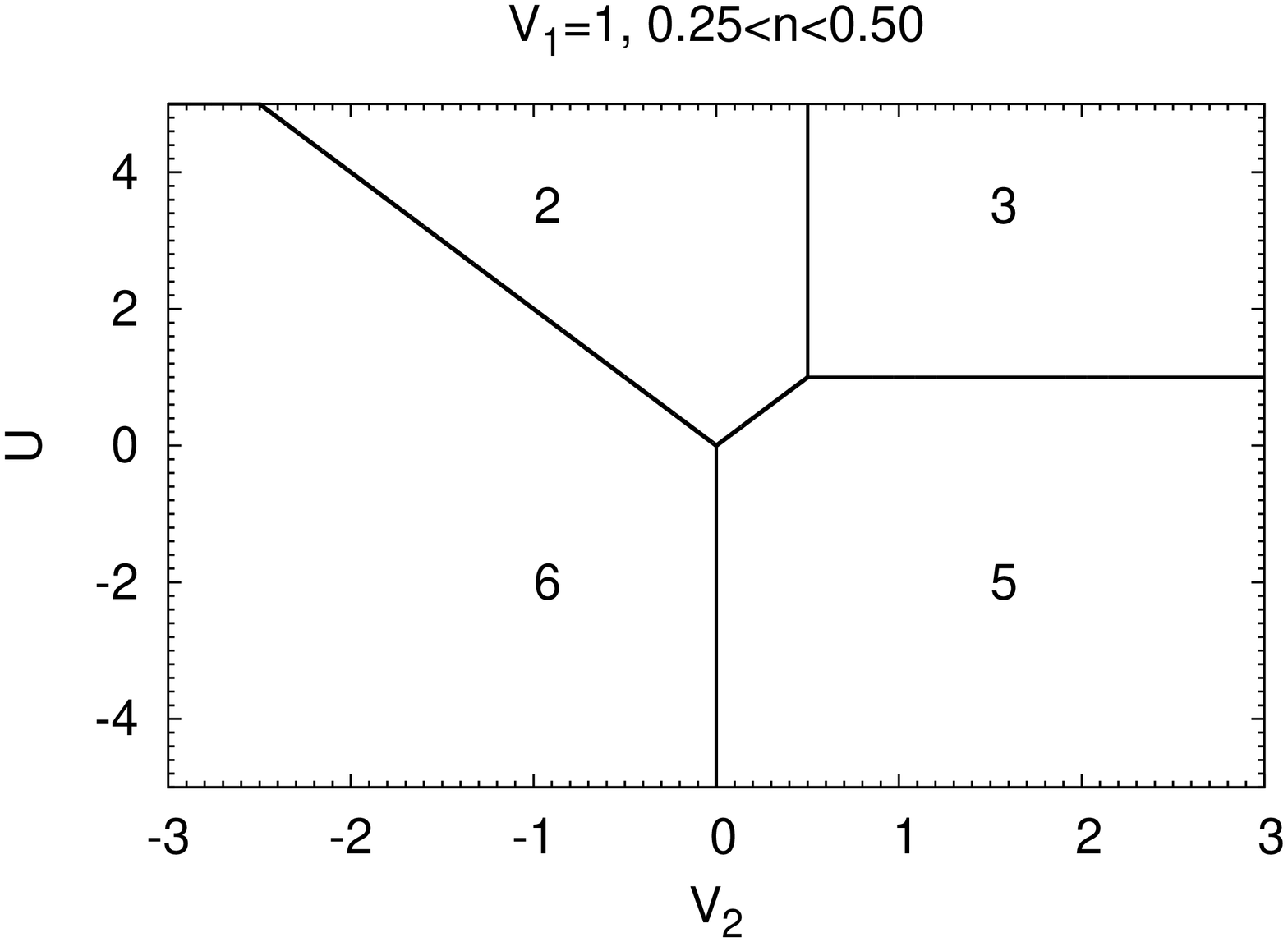}
   \includegraphics[width=4.2cm,angle=270]{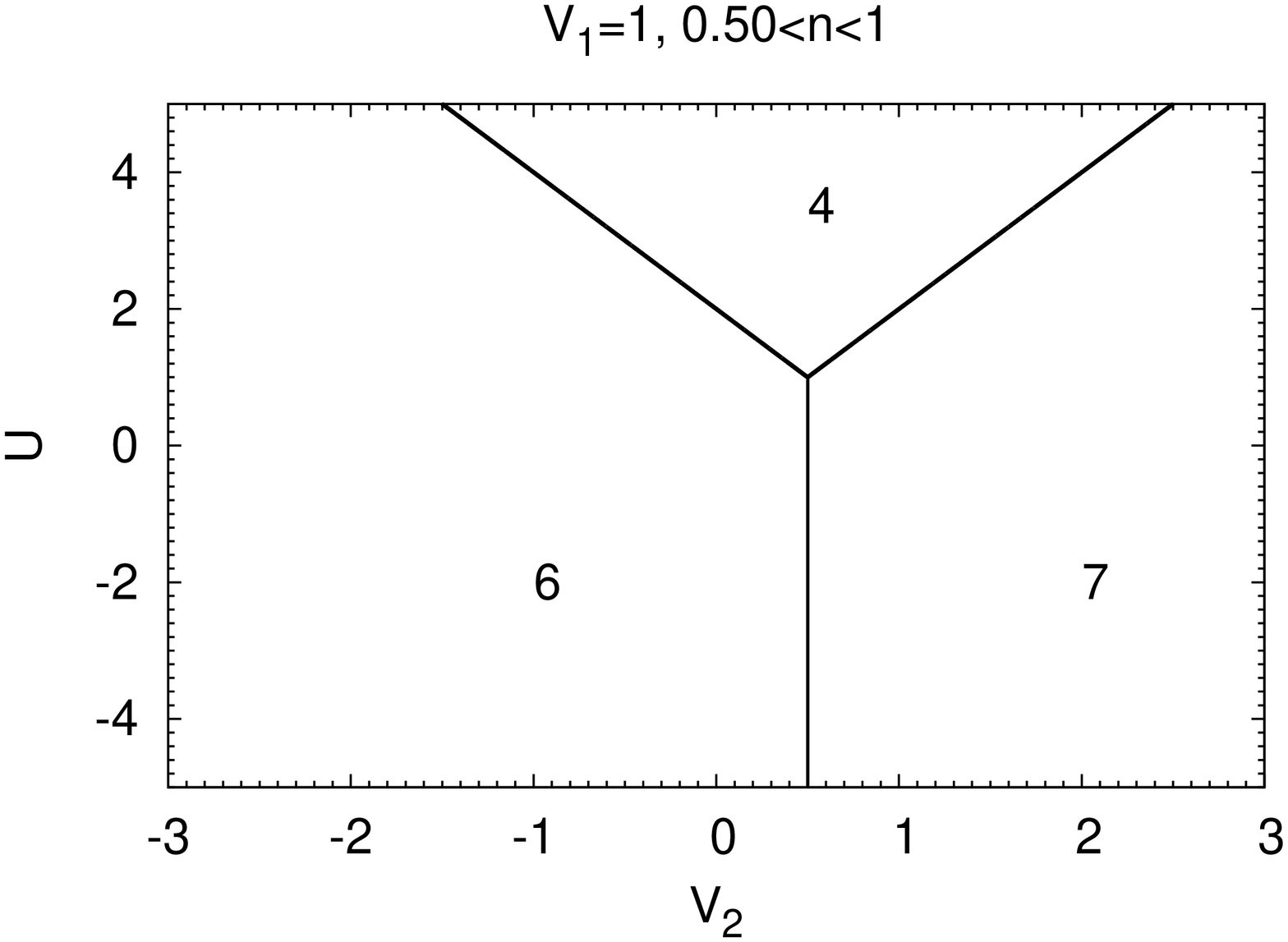}
   \caption{Phase diagram in the case $V_1=1$ and for arbitrary $n$.}
   \label{fig6}
\end{figure*}

On the other hand, the incommensurate macro-phases, defined according to the
criterion explained above, can be divided into three classes depending on the
value of $n$: i) $0<n<\frac{1}{4}$, ii) $\frac{1}{4}<n<\frac{1}{2}$,
iii) $\frac{1}{2}<n<1$. Below we first briefly define the
incommensurate macro-phases, then describe their existence ranges.
The phase diagrams in this case are depicted in Fig.~\ref{fig6}.

\emph{Phase 1:} is identical to the commensurate Phase a.

\emph{Phase 2:} is characterized by the presence of only NNN
correlations ($C_2\ne0$).

\emph{Phase 3:} is characterized by the presence of only NN
correlations ($C_1\ne0$).

\emph{Phase 4:} is characterized by the presence of both NN and
NNN correlations ($C_1\ne0$ and $C_2\ne0$).

\emph{Phase 5:} is characterized by the presence of only double
occupancy correlations ($D\ne0$). The representative pattern for this
phase contains doubly occupied sites, separated by at least two empty
ones.

\emph{Phase 6:} is characterized by the presence of both double
occupancy and NNN correlations ($D\ne0$ and $C_2\ne0$).

\emph{Phase 7:} is characterized by the presence of both double
occupancy and NN correlations ($D\ne0$ and $C_1\ne0$). 

\emph{Phase 8:} is characterized by the presence of all
fundamental correlations: double occupancy, NN and NNN ($D\ne0$
$C_1\ne0$ and $C_2\ne0$).

We proceed now with the description of the phase diagrams at fixed $n$
(incommensurate phase diagrams).
\subsubsection{$0<n<\frac{1}{4}$}
The Phase 1 extends in the range $U>0$ and $V_2>0$. If $V_2$ is large
and negative ($V_2<0$ and $U<-2V_2$) the Phase 6 is stabilized,
while if $U$ is negative ($U<0$ and $V_2>0$) the Phase 5 is
favorite. Finally, when $U>-2V_2$ and $V_2<0$ the Phase 2 realizes the
free energy minimum.
\subsubsection{$\frac{1}{4}<n<\frac{1}{2}$}
Upon increase of doping, the Phase 2 extends towards positive values of
$V_2$ and occupies the region $U>-2V_2$, $U>2V_2$ and $2V_2>V_1$, while
the Phase 5 enlarges to fill the realm $V_2>0$, $U<V_1$ and
$U<2V_1-V_2$. In addition, at $n>\frac{1}{4}$ the Phase 1 is substituted
by the Phase 3 in the range $2V_2>V_1$ and $U>V_1$.
\subsubsection{$\frac{1}{2}<n<1$}
At fillings larger than $\frac{1}{2}$, the Phase 6 extends to $2V_2>V_1$
and $U<2V_1-2V_2$, while two new phases appear: the Phase 7 in the range
$2V_2>V_1$ and $U<2V_2$ and the Phase 4 in the range $U>-2V_2$ and
$U>2V_1-2V_2$.
\begin{figure}
   \centering
   \includegraphics[width=4.2cm,angle=270]{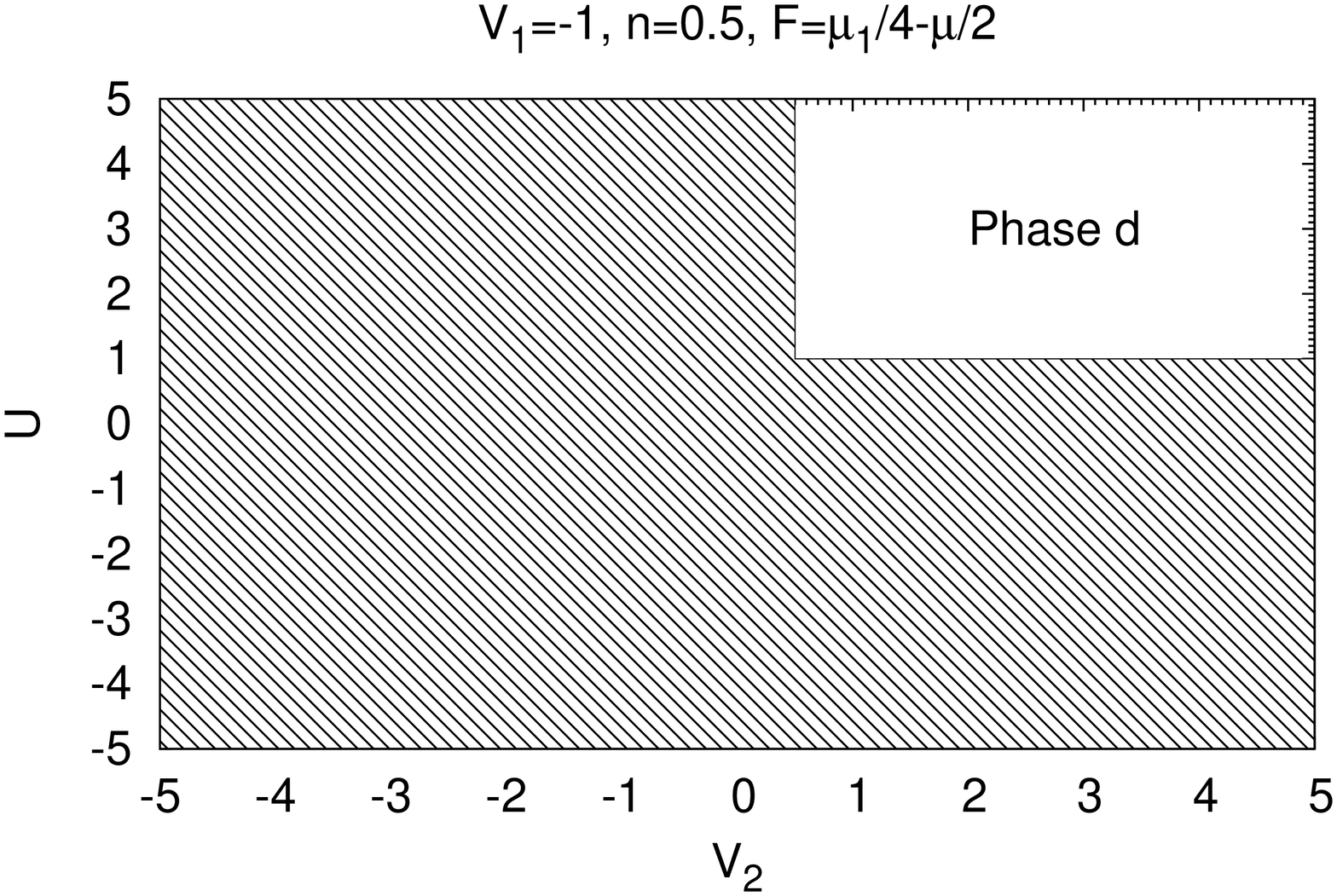}
   \includegraphics[width=4.2cm,angle=270]{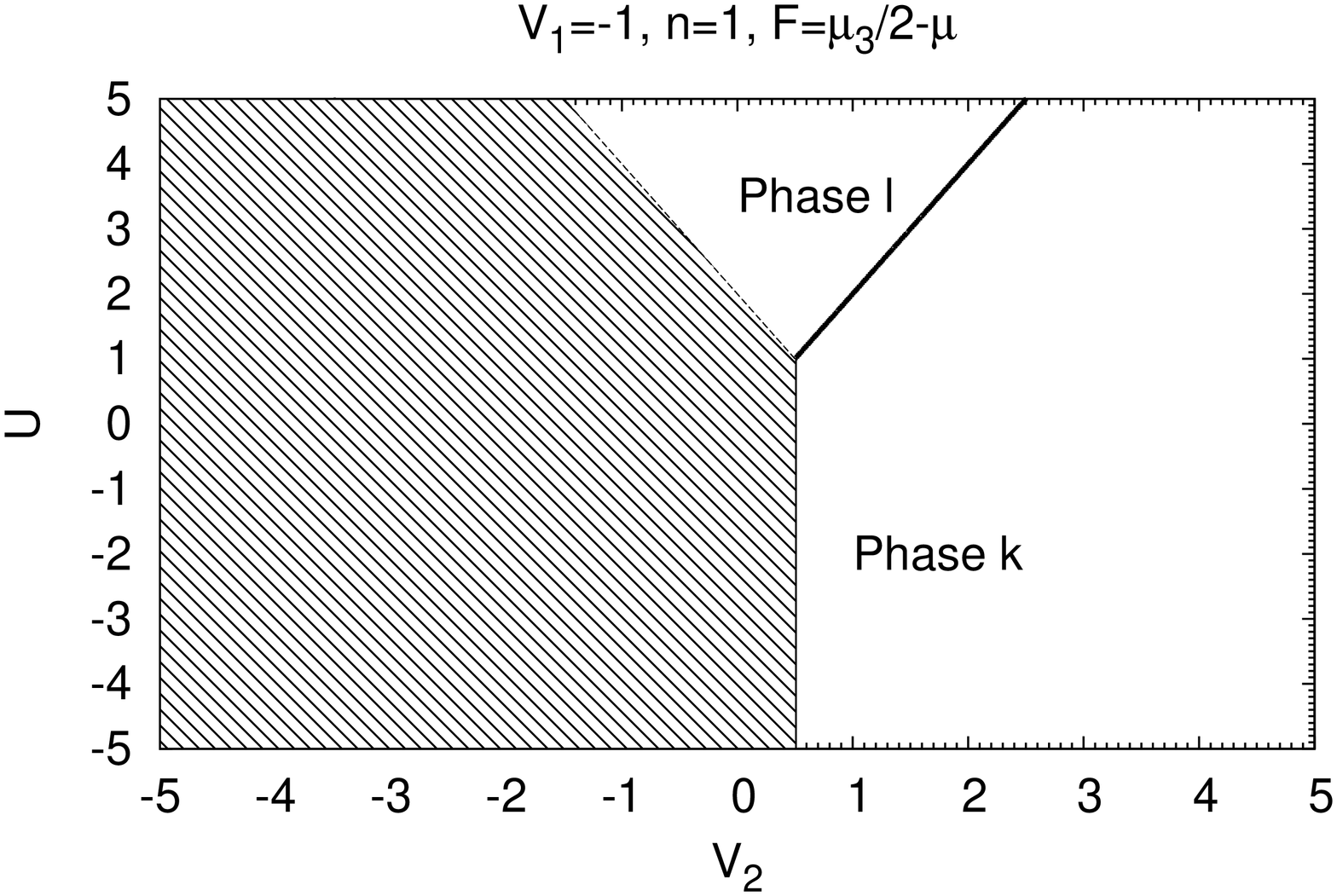}
   \caption{Phase diagram in the case $V_1=-1$ for $n=0.5$ and
	  $n=1$. In the shaded areas the jump in
	  the dependence $\mu(n)$ is absent at the given commensurate $n$.}
   \label{fig4}
\end{figure}

\subsection{Case $V_1=-1$}
Let us consider now the phase diagram in the case $V_1=-1$. The
differences with respect to the case $V_1=1$ emerge when one starts to
consider the existence conditions for the commensurate phases at
$n=\frac{1}{4}$ ($n=\frac{3}{4}$) and $n=\frac{1}{2}$. The former
is equivalent to $\mu_1>0$ and one can easily verify that in our case it
is never satisfied, which means that there are no jumps at
$n=\frac{1}{4}$ and $\frac{3}{4}$ in $\mu(n)$ and the phases at those
precise $n$ do not exist. For the existence of the phase at $n=\frac{1}{2}$ it
is necessary (and sufficient) that $\mu_1<\mu_3$. By inspection, one can
verify that the range of existence of any phase at $n=\frac{1}{2}$ is
$V_2>\frac{1}{2} \land U>1$. Moreover, at $V_1=-1$, the jump at $n=1$
will exist only in the range $V_2>\frac{1}{2}$ and $U>1$, {\it i.e.} in
the same range where the jump at $n=\frac{1}{2}$ exists. In
Fig.~\ref{fig4} we report the phases at $n=\frac{1}{2}$ and $n=1$.
Only the Phases $d$, $l$ and $k$ are present at $V_1=-1$ and their
description has been given above.
%
\begin{figure}
   \centering
   \includegraphics[width=4.2cm,angle=270]{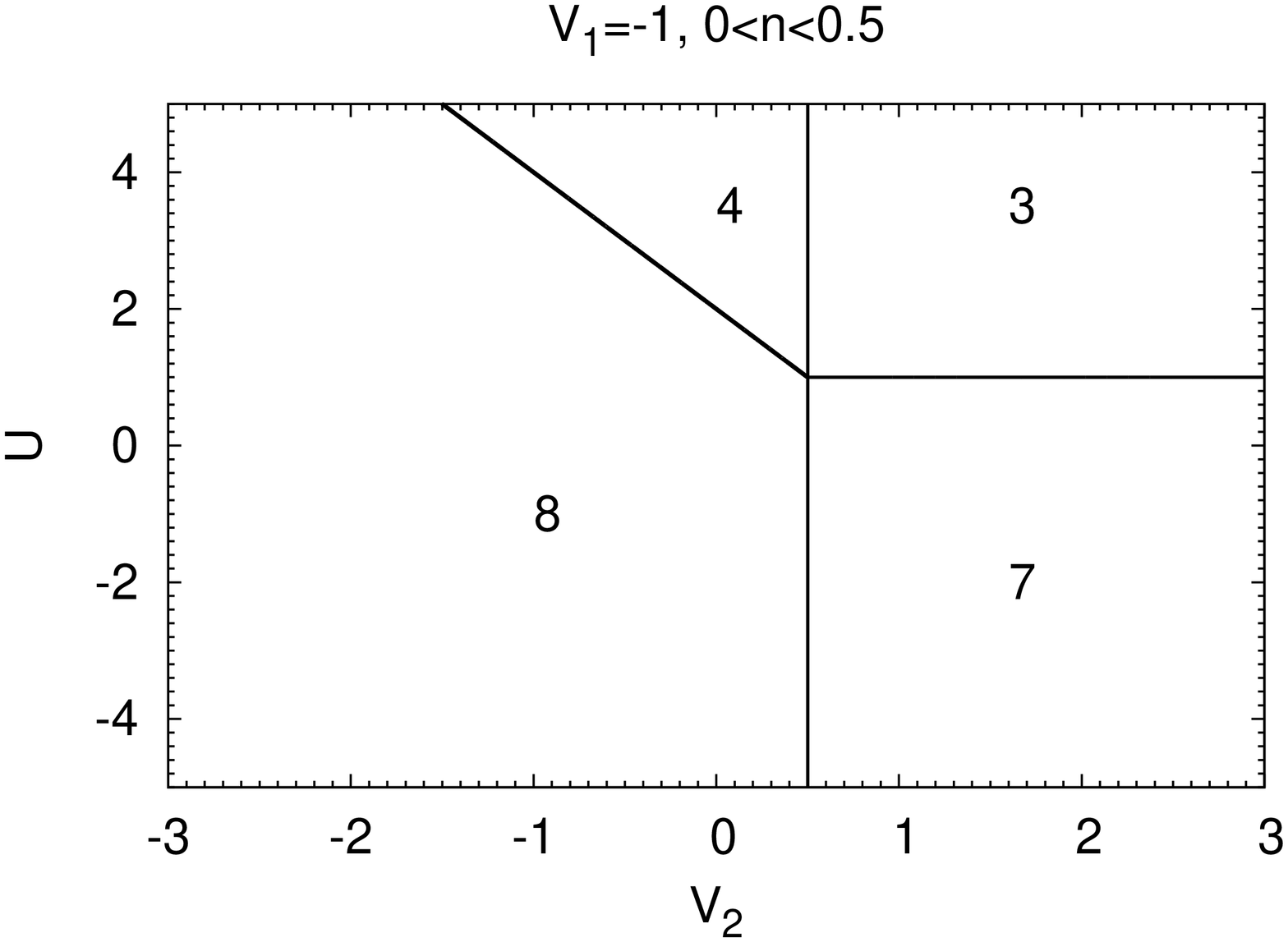}
   \includegraphics[width=4.2cm,angle=270]{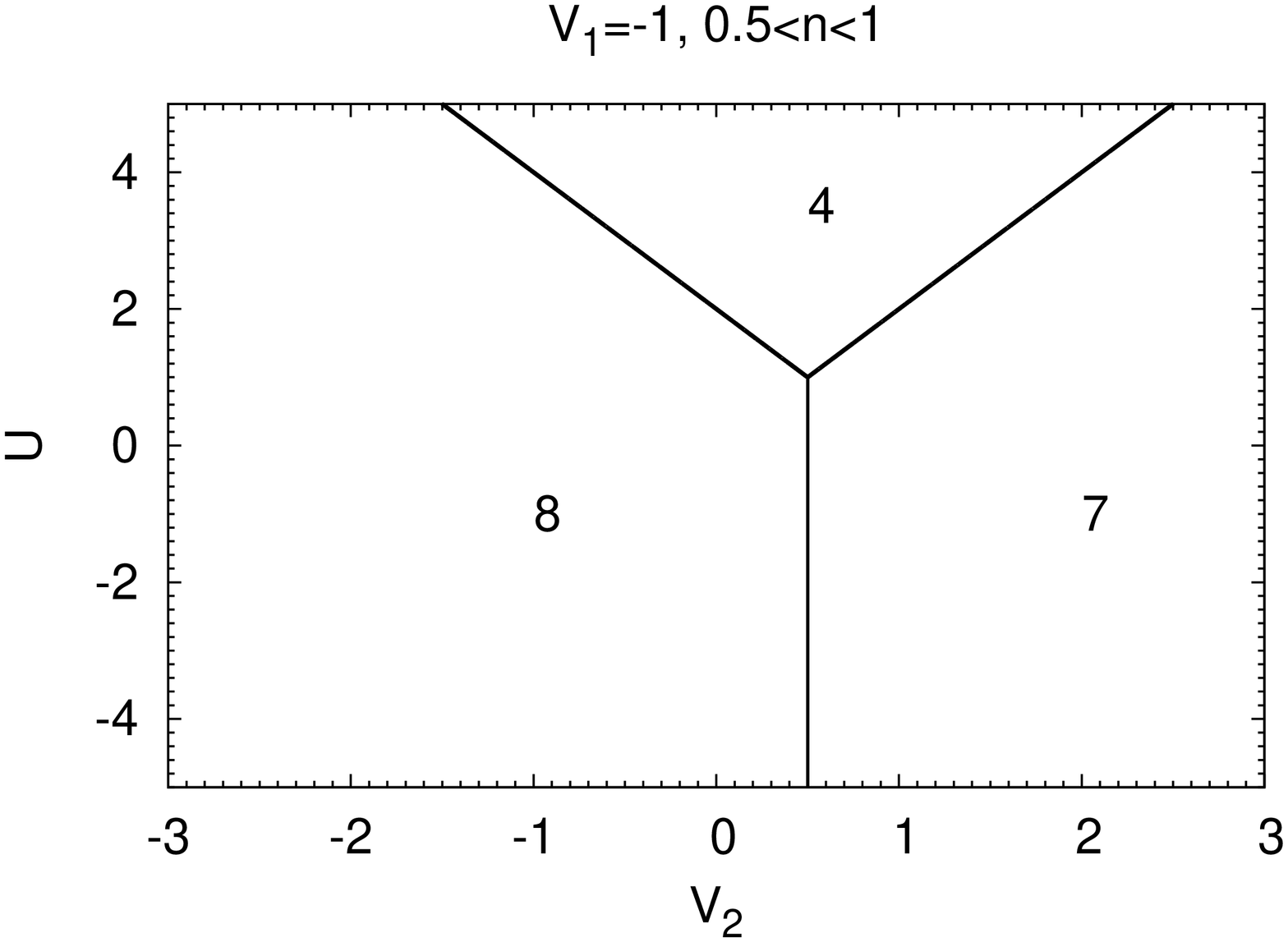}
   \caption{Phase diagram in the case $V_1=-1$ and for arbitrary $n$.
   }
   \label{fig7}
\end{figure}

The incommensurate phase diagram, shown in Fig.~\ref{fig7}, splits into two
cases: $n<\frac{1}{2}$ and $n>\frac{1}{2}$. In addition, only in
this case there appears the Phase 8, in which all the fundamental
CFs are different from zero.
\subsubsection{$0<n<\frac{1}{2}$}
Several phases are present in this case. The Phase 3, characterized by only
NN correlations, extends in the domains: $2V_2>-V_1$, $U>-V_1$; the Phase 4
is placed in the range $2V_2<-V_1$, $U>-2V_1-2V_2$. The Phase 7 exists in the
range $2V_2>-V_1$, $U<-V_1$, while the Phase 8 occupies the area
$\max\left(2V_2, 2V_1-2V_2\right)<U<-2V_1-2V_2$.
\subsubsection{$\frac{1}{2}<n<1$}
At fillings larger than $\frac{1}{2}$, a little changes in the phase
diagram: the range $2V_2>-V_1$, $U>-V_1$, previously dominated by the
Phase 3, is now occupied by the Phases 4 and 7. These latter enlarge and
fill in the following areas: Phase 4 $U>\max\{2V_2, -2V_1-2V_2\}$, Phase 7
$U<2V_2$, $-V_1<2V_2$.
\section{\label{results:therm} Thermodynamic properties}
We have shown in the previous Section that at zero temperature it is
always possible to determine analytically all the properties of the
system at any point in the phase diagram of the Hamiltonian~(\ref{ham}).
At finite $T$, the whole spectrum is involved and the
appearance of various features in the thermodynamic quantities (TQs)
depends on the exact positions
and degeneracies of \textit{all} energy levels in the system.
Therefore, we adopt the following scheme to analyze the low-$T$
properties of TQs: we estimate the level of the charge rigidity of the
ground state, the gaps between the ground state and few low-lying
excited states and the ratio of degeneracy of these excited states with
respect to the ground state degeneracy, when these excited states are well
separated from the rest of the spectrum. Finally, we study the
behavior of the aforementioned TQs across the $T=0$ phase boundaries and
follow the change of the properties upon switching on the temperature.
Following this scheme, we have studied the behaviors of the TQs in
various phases and across various phase boundaries, and report below the
most representative ones. We note that the TQs change not only at the
boundaries of the macro-phases but also crossing the different phases
inside the same macro-phase.
\begin{figure*}
   \centering
   \includegraphics[width=4.2cm,angle=270]{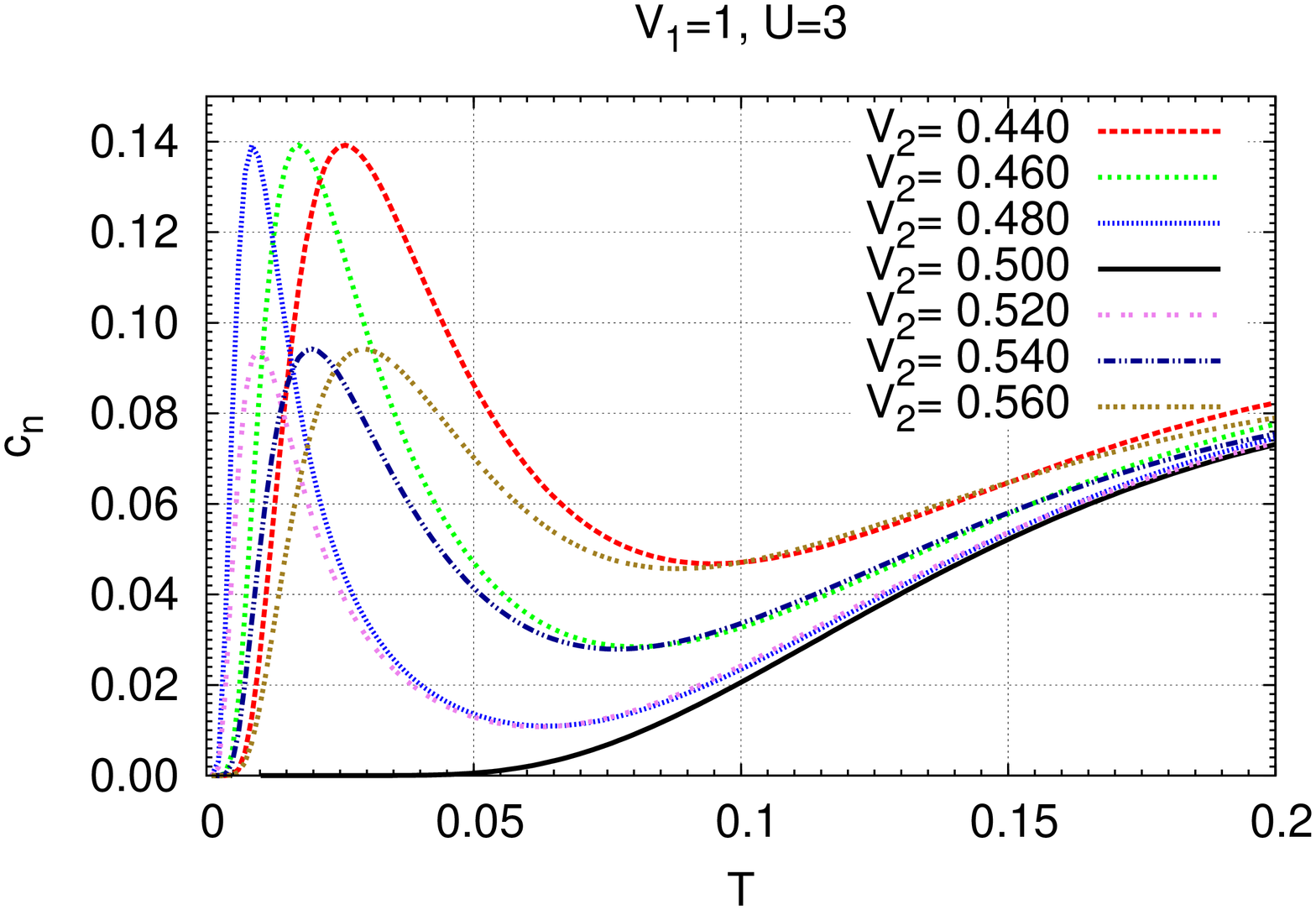}
   \includegraphics[width=4.2cm,angle=270]{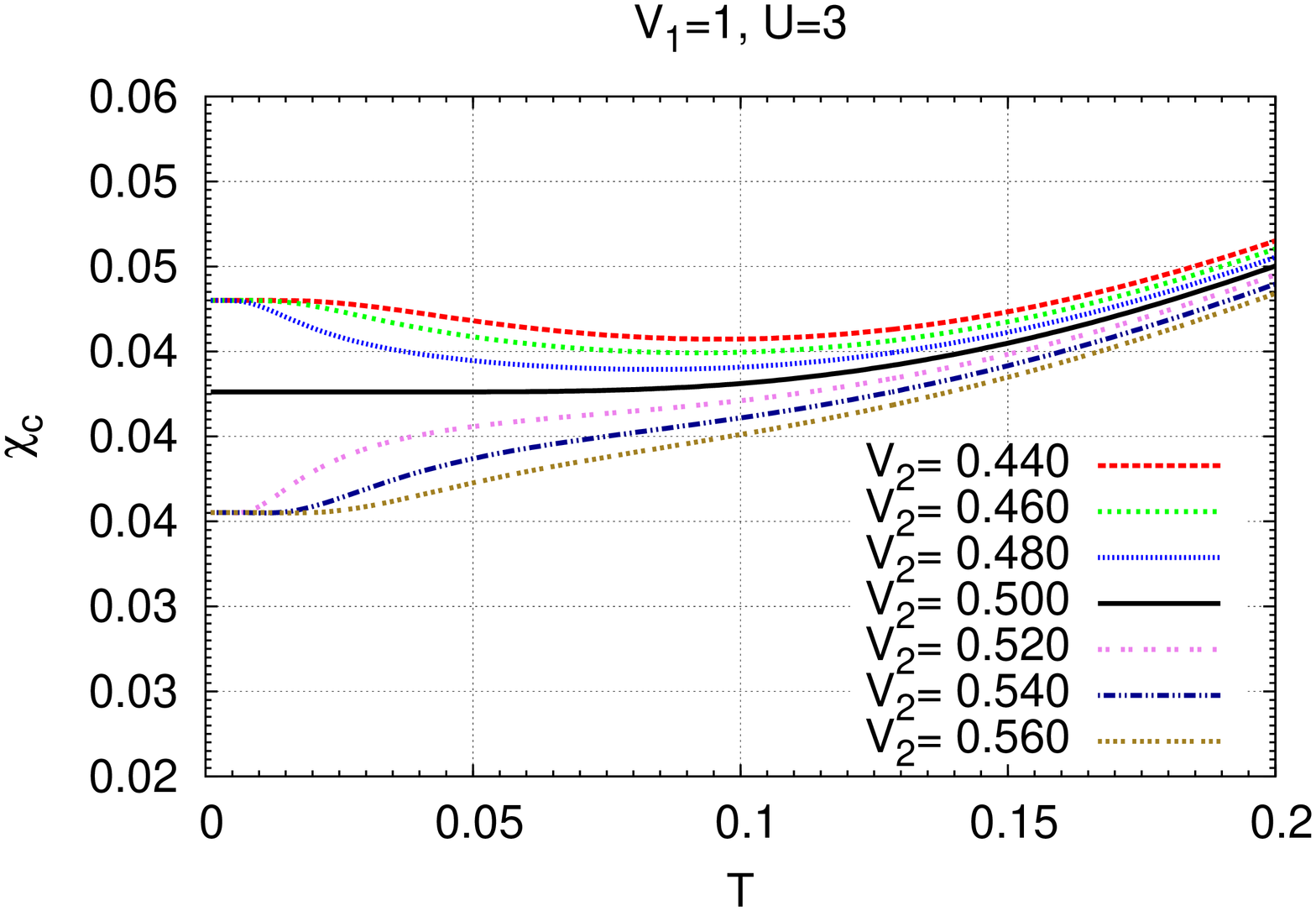}
   \includegraphics[width=4.2cm,angle=270]{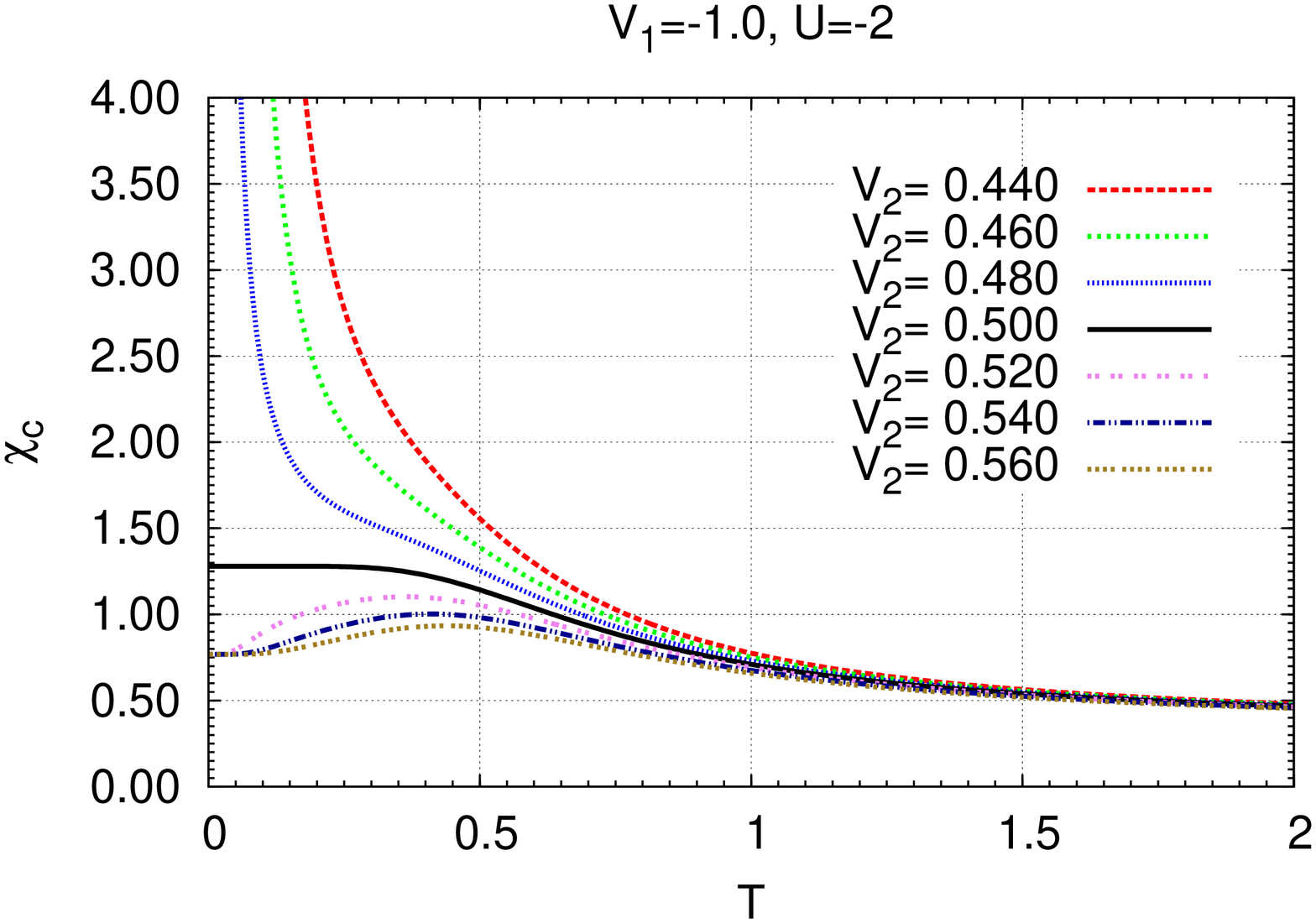}
   \caption{(Color online) Left panel: the behavior of the specific heat $c_n(T)$
   across the phase boundary in a typical $T=0$ phase transition.  $n=0.4$, $V_1=1$,
   $U=3$, the transition occurs at $V_2=0.5$. Middle and right panels:
   the behavior of the charge susceptibility $\chi_c(T)$
   across the phase boundaries in two representative cases. Middle
   panel: $n=0.4$, $V_1=1$, $U=3$, the transition occurs at
   $V_2=0.5$; right panel: $n=0.4$, $V_1=-1$, $U=-2$, the transition
   occurs at $V_2=0.5$.
   }
   \label{fig11}
\end{figure*}

In the present Section, we focus on the following TQs: specific heat,
charge susceptibility and entropy. The specific heat at constant $n$ is
defined as follows:
\be
   c_n = \left(\frac{\partial E}{\partial T}\right)_n,
\ee
where $E$ is the internal energy.
%
%
The specific heat of the model under investigation presents in general a
multiple-peak structure and goes to zero in the limits $T\to 0$
and $T\to \infty$.

We define the charge susceptibility as follows:
\be
\chi_c = 
   T\frac{\partial n}{\partial \mu}.
\ee
$\chi_c$ in the model~(\ref{ham}) can either diverge, go to zero, or go
to a finite limit as $T\to 0$.
A divergent susceptibility signifies the rigidity of the underlying
configuration with respect to the addition of electrons.
\begin{figure}
   \centering
   \includegraphics[width=5.2cm,angle=270]{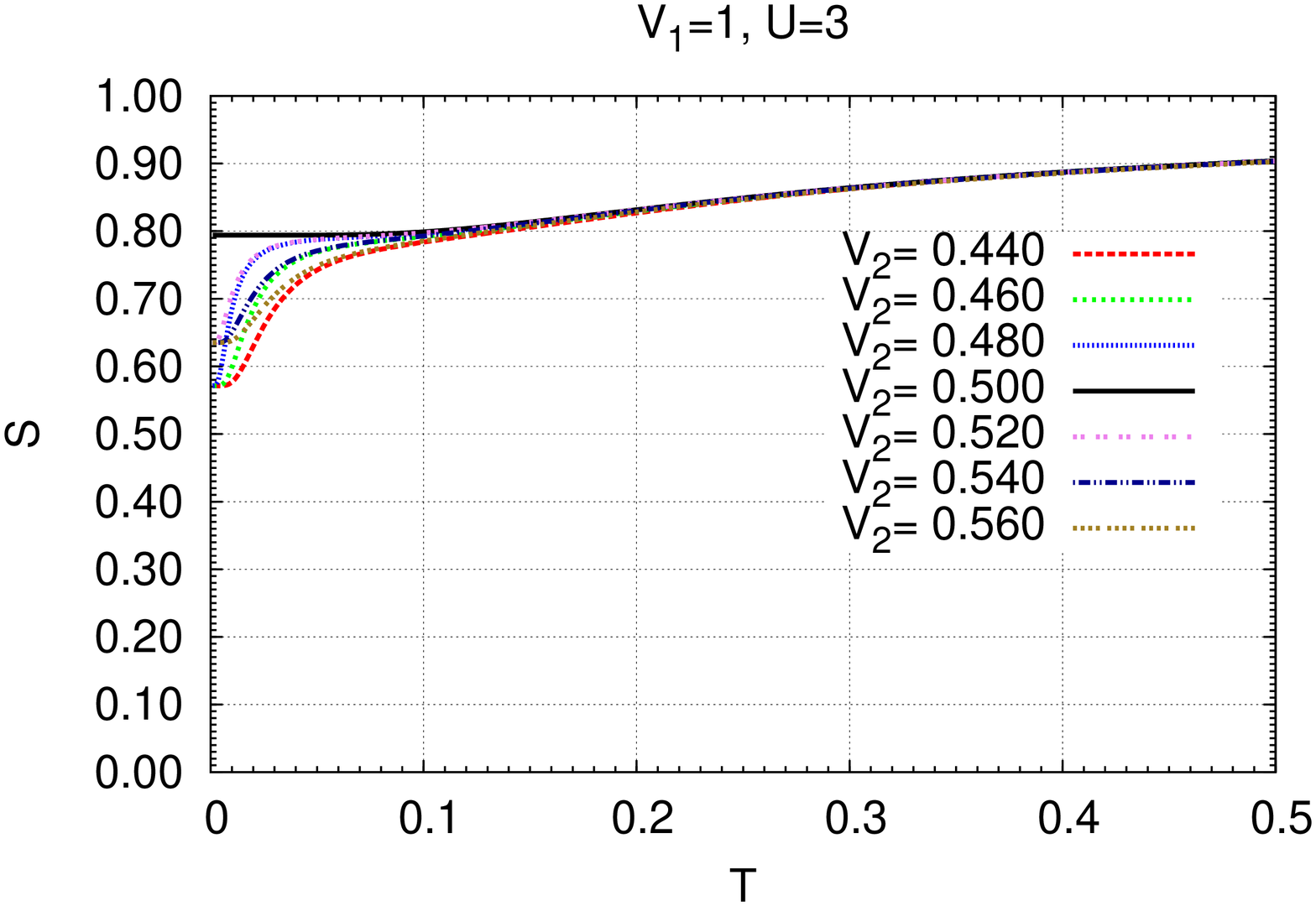}
   \includegraphics[width=5.2cm,angle=270]{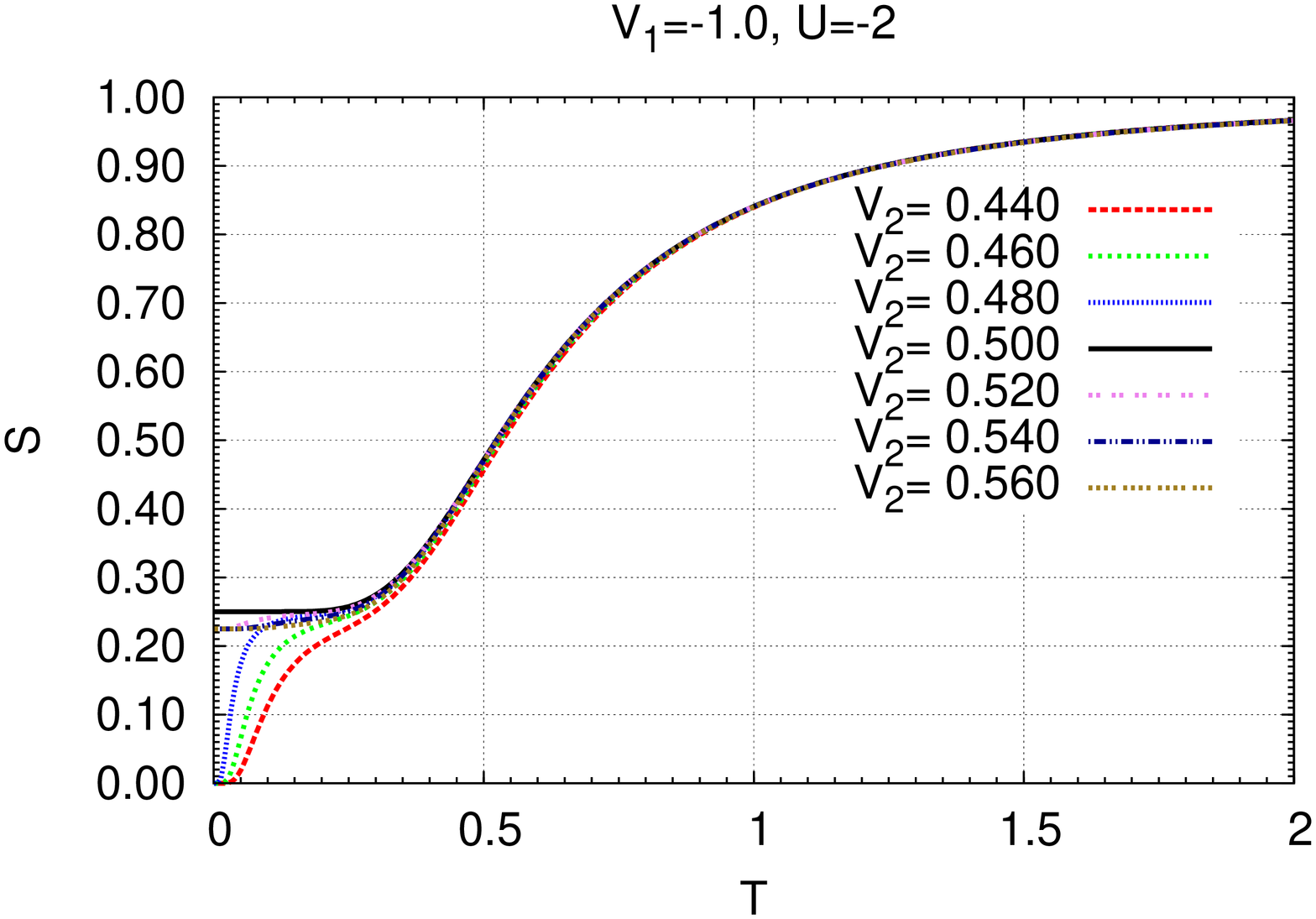}
   \caption{(Color online) The behavior of the entropy $S(T)$ across the phase
   boundaries in two representative cases. Upper panel:
   $n=0.4$, $V_1=1$, $U=3$, the transition occurs at $V_2=0.5$; Lower
   panel: $n=0.4$, $V_1=-1$, $U=-2$, the transition occurs at
   $V_2=0.5$.
   }
   \label{fig13}
\end{figure}

The entropy is defined as the partial derivative of the free
energy with respect to the temperature and, as such, can be easily
computed within the Transfer Matrix method.
One usually
is interested in the $T\to 0$ limit of entropy since it indicates, if
non-zero, a ground-state degeneracy.

In the low-temperature limit, the properties of the system are determined,
to a great extent, by the properties of the ground state and of a few
excited ones. When only the first excited state is involved,
the system resembles a two-level one. We will show in the following that
several features of the two-level system are present in TQs
of~(\ref{ham}) at low temperatures. In order to do that, we
summarize below the results for a generic two-level system. Suppose
that we have a system with two levels $0$ and $\Delta$. Each level can be
degenerate with its own degeneracy $N_0$ and $N_1$, respectively. It can
be easily checked that in such a toy system $c_n$, $S$ and the internal
energy $E$ can be expressed as follows:
\be
   \label{cn_2lev}
   \begin{split}
	  c_n &= \left(\frac{\Delta}{T} \right)^2 \frac{e^{\frac{\Delta}{T}+A}}
	   {\left( 1 + e^{\frac{\Delta}{T}+A} \right)^2}\\
       S   &= \log N_0 +\log\left( 1 + 
	   e^{-\frac{\Delta}{T}-A} \right) +
	   \frac{E}{T}\\
	   E &= \frac{\Delta}{ 1 + e^{\frac{\Delta}{T}+A}},
	\end{split}
\ee
where $A=\log\frac{N_0}{N_1}$.
At fixed $A$, the position of the maximum of $c_n$ depends linearly on $\Delta$:
$T^{\star}=\frac{\Delta}{2\xi}$, where $\xi$ is the solution of the
transcendental equation:
\be
   \xi = \coth\left( \xi + \frac{A}{2} \right),
\ee
while the height of the maximum appears to
be independent of $\Delta$. The limit of entropy at $T\to 0$ is equal to
$\log N_0$ and, thus, is non-zero for degenerate ground states.
After a careful analysis of the whole phase diagram of the model we have
identified the typical behaviors of each TQs across the phase
boundaries.

\textit{Specific heat:} The low-$T$ behavior of the specific heat $c_n$
across the $T=0$ phase transitions is common to the whole phase
diagram. A representative of such behavior is shown in
Fig.~\ref{fig11}. $c_n$ develops a low-$T$ peak, whose position
approaches gradually zero as the system reaches the transition (in case
of Fig.~\ref{fig11} at $V_2=0.5$), while the height of the peak remains
constant. On the opposite side of the transition, the picture is
analogous except for the difference in the peak height.

The appearance of such a low-$T$ peak can be associated to the first
excited state. Indeed, as in the two-level system, the height of the
low-$T$ peak does not depend on the vicinity to the transition $\zeta$,
while the peak position depends linearly on $\zeta$, as can
be seen from Fig.~\ref{fig11}. Here we define $\zeta$ as the shortest
distance from a given point in the phase diagram to the transition line. 
The study of the first excited state not
only sheds light on the low-$T$ behavior of TQs. It appears from our
analysis that the transitions among various $T=0$ phases take place via
the exchange of the ground state with the first excited one. In the
case of Fig.~\ref{fig11}, the transition occurs at $V_2=\frac{1}{2}$;
the ground state energy at $V_2<\frac{1}{2}$ is
$2V_2\left(n-\frac{1}{4}\right)$, the ground state energy at
$V_2>\frac{1}{2}$ is $V_1 \left(n-\frac{1}{4}\right)$, and thus $\Delta=
|V_1-2V_2|\left(n-\frac{1}{4}\right)$. During this transition $V_1=1$
and the path across the transition passes along the line $U=3$.
$\Delta$ decreases linearly as $V_2$ approaches $\frac{1}{2}$ and so
should also do the low-$T$ peak in $c_n$, which is indeed the case, as
shown in Fig.~\ref{fig11}.

\textit{Charge susceptibility and entropy:} Unlikely the specific heat,
charge susceptibility and entropy exhibit several distinct
behaviors: the divergence of $\chi_c$ at $T\to 0$ is accompanied by a
vanishing entropy, while finite $\chi_c$ at $T\to 0$ corresponds to a
finite entropy in the same limit. Examples of such behaviours are shown
Fig.~\ref{fig13}. The divergence of the charge
susceptibility indicates the rigidity of the ground state configuration
with respect to addition of electrons, while a finite entropy at
$T\to 0$ indicates a ground state degeneracy.

\section{\label{conclus} Concluding remarks}

In the present manuscript, we apply the Transfer Matrix method to
find the exact solution for the atomic limit of the 1D Hubbard model
supplied with NN and NNN density-density correlations terms. We utilize
the TM technique in order to fulfil a thorough
analysis of both $T=0$ and finite-temperature properties of the model
under consideration. Several competing interactions induce a quite large
number of $T=0$ phases, and the use of the TM technique
allows us to completely catalogue their properties, including the exact
analytic forms for the internal energy and fundamental correlation
functions. The study of the thermodynamic quantities reveals a few
peculiarities of the system under consideration: a considerable
amount of the phases has a macroscopic degeneracy, manifested by a
finite entropy in the limit $T\to 0$; this is due to the fact that the
Hamiltonian acts only in the charge density sector leaving the
degeneracy with respect to the spin configurations.
On the other hand, even in the charge density sector, a few phases
exhibit a finite entropy at $T\to 0$, owing to the phase separated
states inherent to the atomic models. Some of these states have
been already found in literature in various limiting cases of our
model~\cite{robaszkiewicz_00,mancini_33,Bursill_00}. Physically, such
phases are composed of a superposition of a huge amount of electronic
configurations, each made of blocks of filled sites separated by blocks
of empty ones. Such a block structure gives origin to a large
degeneracy, which often survives in the thermodynamic limit.
Moreover, by studying the low temperature features of the specific heat,
we find that transitions among various phases occur through the
exchange between the first excited and the ground states.

From the viewpoint of methodology, we present a number of useful extensions to the
standard TM technique: symmetrization of the TM, rank reduction of the
TM by use of the system's symmetries, obtaining the ground state phase
diagram from the TM matrix elements and conversion of the grand
canonical phase diagram to the canonical one. Finally, we show how the
properties of the first excited state of the model can be 
inferred from
the low-temperature behavior of the thermodynamic
quantities such as the specific heat, charge susceptibility and entropy.

The study of an easily solvable model such as the one considered here, can be considered as a
starting point for the analysis of more involved models. In particular,
we identify two lines, along which the work is currently in progress: i)
extention of the treatment to an $n$-leg ladder and successively to a 2D
square lattice~\cite{mancini_10}; ii) introduction of the kinetic energy in
order to treat it as a small perturbation.
Plenty of new physical effects should arise upon switching on of the kinetic energy.
In particular, taking into account the extended Coulomb interactions ($V_1$ and $V_2$),
it would be extremely interesting to study the Mott insulating phases at the commensurate fillings
$n_{comm}$. In addition, we expect the spin channel of the model to become active in this case.
\acknowledgments
We acknowledge the CINECA award under the ISCRA initiative (project
MP34dTMO), for the availability of high performance computing resources
and support.

\appendix

\section{\label{app:a}Explicit form of the TM}
For the sake of completeness, we report here the explicit form of the TM for the model~(\ref{ham}).
Here the quantities $F_{n}^i$ denote the exponentiated energy scales from Table~\ref{en_scales} ($E^i_n$)
according to the following definition:
\be
   F^i_{n}=\exp\left( -\frac{E_{n}^i}{T}\right),
\ee
where the lower index $n$ denotes one of the commensurate fillings $n_{comm}$.
If several energy scales correspond to a given
$n$, these are labeled by the upper index $i$ corresponding to the order in which the
energy scales appear in Table~\ref{en_scales}.
\begin{widetext}
\be
Z(1,2)=\left(
\begin{array}{cccccccccccccccc}
 1 & F_{\frac{1}{4}} & F_{\frac{1}{4}} & F_{\frac{1}{2}}^2 & F_{\frac{1}{4}} & F_{\frac{1}{2}}^3 & F_{\frac{1}{2}}^3 & F_{\frac{3}{4}}^2 &
   F_{\frac{1}{4}} & F_{\frac{1}{2}}^3 & F_{\frac{1}{2}}^3 & F_{\frac{3}{4}}^2 & F_{\frac{1}{2}}^2 & F_{\frac{3}{4}}^2 & F_{\frac{3}{4}}^2 &
   F_1^2 \\
 F_{\frac{1}{4}} & F_{\frac{1}{2}}^1 & F_{\frac{1}{2}}^1 & F_{\frac{3}{4}}^1 & F_{\frac{1}{2}}^3 & F_{\frac{3}{4}}^3 & F_{\frac{3}{4}}^3 & F_1^3 &
   F_{\frac{1}{2}}^3 & F_{\frac{3}{4}}^3 & F_{\frac{3}{4}}^3 & F_1^3 & F_{\frac{3}{4}}^2 & F_1^4 & F_1^4 & F_{\frac{5}{4}}^2 \\
 F_{\frac{1}{4}} & F_{\frac{1}{2}}^1 & F_{\frac{1}{2}}^1 & F_{\frac{3}{4}}^1 & F_{\frac{1}{2}}^3 & F_{\frac{3}{4}}^3 & F_{\frac{3}{4}}^3 & F_1^3 &
   F_{\frac{1}{2}}^3 & F_{\frac{3}{4}}^3 & F_{\frac{3}{4}}^3 & F_1^3 & F_{\frac{3}{4}}^2 & F_1^4 & F_1^4 & F_{\frac{5}{4}}^2 \\
 F_{\frac{1}{2}}^2 & F_{\frac{3}{4}}^1 & F_{\frac{3}{4}}^1 & F_1^1 & F_{\frac{3}{4}}^2 & F_1^3 & F_1^3 &
F_{\frac{5}{4}}^1 & F_{\frac{3}{4}}^2 & F_1^3
   & F_1^3 & F_{\frac{5}{4}}^1 & F_1^2 & F_{\frac{5}{4}}^2 & F_{\frac{5}{4}}^2 & F_{\frac{3}{2}}^2 \\
 F_{\frac{1}{4}} & F_{\frac{1}{2}}^3 & F_{\frac{1}{2}}^3 & F_{\frac{3}{4}}^2 & F_{\frac{1}{2}}^1 & F_{\frac{3}{4}}^3 & F_{\frac{3}{4}}^3 & F_1^4
   & F_{\frac{1}{2}}^1 & F_{\frac{3}{4}}^3 & F_{\frac{3}{4}}^3 & F_1^4 & F_{\frac{3}{4}}^1 & F_1^3 & F_1^3 & F_{\frac{5}{4}}^2 \\
 F_{\frac{1}{2}}^3 & F_{\frac{3}{4}}^3 & F_{\frac{3}{4}}^3 & F_1^3 & F_{\frac{3}{4}}^3 & F_1^5 & F_1^5 & F_{\frac{5}{4}}^3 & F_{\frac{3}{4}}^3
   & F_1^5 & F_1^5 & F_{\frac{5}{4}}^3 & F_1^3 & F_{\frac{5}{4}}^3 & F_{\frac{5}{4}}^3 & F_{\frac{3}{2}}^3 \\
 F_{\frac{1}{2}}^3 & F_{\frac{3}{4}}^3 & F_{\frac{3}{4}}^3 & F_1^3 & F_{\frac{3}{4}}^3 & F_1^5 & F_1^5 & F_{\frac{5}{4}}^3 & F_{\frac{3}{4}}^3
   & F_1^5 & F_1^5 & F_{\frac{5}{4}}^3 & F_1^3 & F_{\frac{5}{4}}^3 & F_{\frac{5}{4}}^3 & F_{\frac{3}{2}}^3 \\
 F_{\frac{3}{4}}^2 & F_1^3 & F_1^3 & F_{\frac{5}{4}}^1 & F_1^4 & F_{\frac{5}{4}}^3 &
F_{\frac{5}{4}}^3 & F_{\frac{3}{2}}^1 & F_1^4 &
   F_{\frac{5}{4}}^3 & F_{\frac{5}{4}}^3 & F_{\frac{3}{2}}^1 & F_{\frac{5}{4}}^2 & F_{\frac{3}{2}}^3 & F_{\frac{3}{2}}^3 & F_{\frac{7}{4}} \\
 F_{\frac{1}{4}} & F_{\frac{1}{2}}^3 & F_{\frac{1}{2}}^3 & F_{\frac{3}{4}}^2 & F_{\frac{1}{2}}^1 & F_{\frac{3}{4}}^3 & F_{\frac{3}{4}}^3 & F_1^4
   & F_{\frac{1}{2}}^1 & F_{\frac{3}{4}}^3 & F_{\frac{3}{4}}^3 & F_1^4 & F_{\frac{3}{4}}^1 & F_1^3 & F_1^3 & F_{\frac{5}{4}}^2 \\
 F_{\frac{1}{2}}^3 & F_{\frac{3}{4}}^3 & F_{\frac{3}{4}}^3 & F_1^3 & F_{\frac{3}{4}}^3 & F_1^5 & F_1^5 & F_{\frac{5}{4}}^3 & F_{\frac{3}{4}}^3
   & F_1^5 & F_1^5 & F_{\frac{5}{4}}^3 & F_1^3 & F_{\frac{5}{4}}^3 & F_{\frac{5}{4}}^3 & F_{\frac{3}{2}}^3 \\
 F_{\frac{1}{2}}^3 & F_{\frac{3}{4}}^3 & F_{\frac{3}{4}}^3 & F_1^3 & F_{\frac{3}{4}}^3 & F_1^5 & F_1^5 & F_{\frac{5}{4}}^3 & F_{\frac{3}{4}}^3
   & F_1^5 & F_1^5 & F_{\frac{5}{4}}^3 & F_1^3 & F_{\frac{5}{4}}^3 & F_{\frac{5}{4}}^3 & F_{\frac{3}{2}}^3 \\
 F_{\frac{3}{4}}^2 & F_1^3 & F_1^3 & F_{\frac{5}{4}}^1 & F_1^4 & F_{\frac{5}{4}}^3 &
F_{\frac{5}{4}}^3 & F_{\frac{3}{2}}^1 & F_1^4 &
   F_{\frac{5}{4}}^3 & F_{\frac{5}{4}}^3 & F_{\frac{3}{2}}^1 & F_{\frac{5}{4}}^2 & F_{\frac{3}{2}}^3 & F_{\frac{3}{2}}^3 & F_{\frac{7}{4}} \\
 F_{\frac{1}{2}}^2 & F_{\frac{3}{4}}^2 & F_{\frac{3}{4}}^2 & F_1^2 & F_{\frac{3}{4}}^1 & F_1^3 & F_1^3
& F_{\frac{5}{4}}^2 & F_{\frac{3}{4}}^1 &
   F_1^3 & F_1^3 & F_{\frac{5}{4}}^2 & F_1^1 & F_{\frac{5}{4}}^1 & F_{\frac{5}{4}}^1 & F_{\frac{3}{2}}^2 \\
 F_{\frac{3}{4}}^2 & F_1^4 & F_1^4 & F_{\frac{5}{4}}^2 & F_1^3 & F_{\frac{5}{4}}^3 & F_{\frac{5}{4}}^3 & F_{\frac{3}{2}}^3 & F_1^3 &
   F_{\frac{5}{4}}^3 & F_{\frac{5}{4}}^3 & F_{\frac{3}{2}}^3 & F_{\frac{5}{4}}^1 & F_{\frac{3}{2}}^1
& F_{\frac{3}{2}}^1 & F_{\frac{7}{4}} \\
 F_{\frac{3}{4}}^2 & F_1^4 & F_1^4 & F_{\frac{5}{4}}^2 & F_1^3 & F_{\frac{5}{4}}^3 & F_{\frac{5}{4}}^3 & F_{\frac{3}{2}}^3 & F_1^3 &
   F_{\frac{5}{4}}^3 & F_{\frac{5}{4}}^3 & F_{\frac{3}{2}}^3 & F_{\frac{5}{4}}^1 & F_{\frac{3}{2}}^1
& F_{\frac{3}{2}}^1 & F_{\frac{7}{4}} \\
 F_1^2 & F_{\frac{5}{4}}^2 & F_{\frac{5}{4}}^2 & F_{\frac{3}{2}}^2 & F_{\frac{5}{4}}^2 & F_{\frac{3}{2}}^3 & F_{\frac{3}{2}}^3 &
   F_{\frac{7}{4}} & F_{\frac{5}{4}}^2 & F_{\frac{3}{2}}^3 & F_{\frac{3}{2}}^3 & F_{\frac{7}{4}} & F_{\frac{3}{2}}^2 & F_{\frac{7}{4}} &
   F_{\frac{7}{4}} & F_2 \\
\end{array}
\right).
\ee
\end{widetext}
\begin{table*}
\begin{ruledtabular}
   \caption{\label{tab_mphases}
   The complete list of the phases of the model~(\ref{ham}), their
   existence ranges and ground-state energies.}
   \begin{tabular}{l|l|l|l}
	  Phase & Existence range $V_2$, $U$ & Existence range $n$ & Energy per site \\
   \hline \hline
   \multicolumn{4}{c}{$V_1=+1$} \\
   \hline \hline
   I    & $U>0 \land V_2>0$     & $0< n < \frac{1}{4}$ & $0$ \\ 
   \hline
   II   & $U<-2V_2 \land V_2<0$ & $0< n< 1$            & $U\frac{n}{2}+2V_2 n$ \\
   \hline
   III  & $U>-2V_2 \land V_2<0$ & $0< n < \frac{1}{2}$ & $V_2 n$\\
   IV   & $U<0 \land V_2>0$     & $0< n < \frac{1}{2}$ & $\frac{Un}{2}$ \\
   \hline
   V    & $U>2V_2 \land 0<V_2<\frac{1}{2}$ & $\frac{1}{4} < n < \frac{1}{2}$ & $2V_2 \left(n - \frac{1}{4} \right)$ \\
   VI   & $U>1 \land V_2>\frac{1}{2}$      & $\frac{1}{4} < n < \frac{1}{2}$ & $ V_1 \left(n - \frac{1}{4} \right)$ \\
   VII  & $0<U<1 \land 2V_2>U$             & $\frac{1}{4} < n < \frac{1}{2}$ & $   U \left(n - \frac{1}{4} \right)$\\
   \hline
   VIII & $-2V_2<U<2-2V_2 \land V_2<0$     & $\frac{1}{2} < n < 1$           & $U \left(n - \frac{1}{2}\right) + V_2(3n-1)$\\
   IX   & $U>2-2V_2 \land V_2<0$           & $\frac{1}{2} < n < 1$           & $V_1\left(2n - 1 \right) + V_2 n$\\
   \hline
   X   & $U>-2V_2 \land 0<V_2<\frac{1}{2}$& $\frac{1}{2} < n < \frac{3}{4}$ & $V_1\left(2n - 1\right) + \frac{V_2}{2}$ \\
   XI  & $U>2V_2 \land V_2>\frac{1}{2}$   & $\frac{1}{2} < n < \frac{3}{4}$ & $V_2\left(2n - 1 \right)+
   V_1\left(n-\frac{1}{4}\right)$\\
   XII & $1<U<2V_2$                       & $\frac{1}{2} < n < \frac{3}{4}$ & $U\left(n - \frac{1}{2} \right) + V_1 
   \left( n - \frac{1}{4} \right)$ \\
   XIII & $2V_2<U<2-2V_2 \land 0<V_2<\frac{1}{2}$ & $\frac{1}{2} < n < \frac{3}{4}$ & $U\left(n - \frac{1}{2} \right)
   + 2V_2 \left( n - \frac{1}{4} \right)$\\
   XIV  & $0<U<2V_2 \land 0<V_2<\frac{1}{2}$      & $\frac{1}{2} < n < \frac{3}{4}$ & $2V_2\left(2n - 1\right) + \frac{U}{4}$\\
   XV   & $0<U<1 \land V_2>\frac{1}{2}$           & $\frac{1}{2} < n < \frac{3}{4}$ & $V_1 \left(2n - 1\right) + \frac{U}{4}$\\
   \hline
   XVI  & $U<0 \land V_2>\frac{1}{2}$             & $\frac{1}{2}< n < 1$            & $V_1 \left(2n - 1\right) + \frac{Un}{2}$\\
   XVII & $U<0 \land 0<V_2<\frac{1}{2}$           & $\frac{1}{2}< n < 1$            & $2V_2\left(2n - 1\right) + \frac{Un}{2}$\\
   XVIII& $U>\max\{2V_2, 2-2V_2\} \land V_2>0$    & $\frac{3}{4} < n < 1$           & $2(V_1+V_2)\left(n - \frac{1}{2}\right)$ \\
   XIX  & $U<2V_2 \land V_2>\frac{1}{2}$          & $\frac{3}{4} < n < 1$           & $(U+2V_1)\left(n - \frac{1}{2}\right)$\\
   XX   & $0<U<2-2V_2 \land 0<V_2<\frac{1}{2}$    & $\frac{3}{4} < n < 1$           & $(U+4V_2)\left(n - \frac{1}{2}\right)$ \\
   \hline \hline
   \multicolumn{4}{c}{$V_1=-1$} \\
   \hline \hline
   XXI  & $U<2-2V_2 \land V_2<\frac{1}{2}$        & $0< n< 2$                       & $(U+4V_1+4V_2) \frac{n}{2}$\\
   XXII & $U>2-2V_2 \land V_2<\frac{1}{2}$        & $0< n< 1$                       & $(V_1+V_2 ) n$\\
   XXIII&$U<1 \land V_2>\frac{1}{2}$             & $0< n< 1$                       & $\left(\frac{U}{2}+V_1\right)n$\\
   \hline
   XXIV &$U>1       \land V_2>\frac{1}{2}$        & $0< n< \frac{1}{2}$             & $\frac{V_1n}{2}$ \\
   \hline
   XXV   &$U>2V_2 \land V_2>\frac{1}{2}$          & $\frac{1}{2}< n< 1$             & $\frac{V_1}{2}\left( 3n-1\right) + 2V_2 (n-\frac{1}{2})$ \\
   XXVI  &$1<U<2V_2 \land V_2>\frac{1}{2}$        & $\frac{1}{2}< n< 1$         & $\frac{V_1}{2}\left( 3n-1\right) + U (n-\frac{1}{2})$ \\
\end{tabular}
\end{ruledtabular}
\end{table*}
\section{\label{app:e}Micro-phases}
In Table~\ref{tab_mphases}, we present all the micro-phases of the model under investigation and
summarize their properties. Namely, for each micro-phase we report the existence ranges both in
terms of $U, V_1, V_2$ and in terms of $n$ as well as the exact expression for the ground state energy,
which can be used to extract the fundamental correlation functions.
\bibliographystyle{apsrev4-1}
\bibliography{article}
\end{document}